\documentclass[a4paper]{aa}
\usepackage[english,english]{babel}
\usepackage{amsmath}
\usepackage{amssymb,amsfonts,textcomp}
\usepackage{array}
\usepackage{supertabular}
\usepackage{hhline}
\usepackage{hyperref}
\usepackage[usenames]{color}

\usepackage{xspace}
\usepackage{graphicx}
\usepackage{color}
\usepackage{txfonts}
\usepackage{natbib}
\usepackage{lscape}
\usepackage{longtable}
\usepackage{placeins}
\usepackage{paralist}
\usepackage{balance}

\def\gtrsim{\lower.5ex\hbox{$\; \buildrel > \over \sim \;$}}
\usepackage{graphicx}

\newcommand{\der}{{\rm d}}
\newcommand{\dd}{{\rm d}}
\newcommand{\be}{\begin{equation}}
\newcommand{\ee}{\end{equation}}

\renewcommand{\vec}[1]{\mathbf #1}

\newmuskip\pFqskip
\mathchardef\pFcomma=\mathcode`, 

\newcommand*\pFq[5]{%
  \begingroup
  \begingroup\lccode`~=`,
    \lowercase{\endgroup\def~}{\pFcomma\mkern\pFqskip}%
  \mathcode`,=\string"8000
  {}_{#1}F_{#2}\left(#3;#4;#5\right)%
  \endgroup
}

\def \apjl {{\it ApJ Let.}}
\def \apjs {{\it ApJ Sup.}}

\def \aap {{\it A\&A}}

\begin{document}
\title{On the projected mass distribution around galaxy clusters : 
}
\subtitle{a Lagrangian theory of harmonic power spectra}

\author{
Sandrine Codis \inst{1}\thanks{E-mail:~\tt{codis@cita.utoronto.ca}}
  \and Rapha\"el Gavazzi\inst{2} \and
  Christophe Pichon\inst{2,3}{ \and C\'eline Gouin\inst{2}}
}
\institute{
Canadian Institute for Theoretical Astrophysics, University of Toronto, 60 St. George Street, Toronto, ON M5S 3H8, Canada
\label{inst1}
\and
Institut d'Astrophysique de Paris, UMR7095 CNRS \& Universit\'e Pierre et Marie Curie, 98bis Bd Arago, F-75014, Paris, France
\label{inst2}
\and
Korea Institute of Advanced Studies (KIAS) 85 Hoegiro, Dongdaemun-gu, Seoul, 02455, Republic of Korea
\label{inst3}
}

\date{\today}


\abstract
{}
{
Gravitational lensing allows to quantify the angular distribution of the convergence field around clusters of galaxies to constrain their
connectivity to the cosmic web. 
We describe in this paper the corresponding theory in Lagrangian space where analytical results can be obtained by identifying clusters to peaks in the initial field.
}
{
We derive the three-point Gaussian statistics of a two-dimensional field and its first and second derivatives. 
The formalism allows us to study the statistics of the field in a shell around a central peak, in particular its multipolar decomposition.
}
{
The peak condition is shown to significantly remove power from the dipolar contribution and to modify the monopole and quadrupole. As expected, higher order multipoles are not significantly modified by the constraint. Analytical predictions are successfully checked against measurements in Gaussian random fields. The effect of substructures and radial weighting is shown to be small and does not change the qualitative picture. { {The non-linear evolution is shown to induce a non-linear bias of all multipoles proportional to the cluster mass.}}
}
{We predict the Gaussian { {and weakly non-Gaussian} }statistics of multipolar moments of a two-dimensional field around a peak as a proxy for the azimuthal distribution of the convergence field around a cluster of galaxies.  A quantitative estimate of this multipolar decomposition of the convergence field around clusters in numerical simulations of structure formation and in observations will be presented in two forthcoming papers.}
   
\authorrunning{S.~Codis et al.}
\titlerunning{Projected mass distribution around clusters}

\maketitle

\section{Introduction}

Galaxies are not islands uniformly distributed in the Universe.
Over the last decades and with the increasing precision of both observations and simulations, they have been shown to reside in a complex network made of large filaments surrounded by walls and voids and intersecting at the overdense nodes of this so-called cosmic web \citep{ks93,bkp96}. From the pioneering works of Zeldovich in the seventies  to the peak-patch picture of \cite{bm96a}, the anisotropic nature of the gravitational collapse have been used to explain the birth and growth of the cosmic web. The origin of filaments and nodes lies in the asymmetries of the initial Gaussian random field describing the primordial universe and amplified by gravitational collapse. The above-mentioned works pointed out the importance of non-local tidal effects in weaving the cosmic web. The high-density peaks define the nodes of the evolving cosmic web and completely determine the filamentary pattern in between. In particular, one can appreciate the crucial role played by the study of constrained random fields in understanding the geometry of the large-scale matter distribution.

Galaxy clusters sitting at these nodes are continuously fed by their connected filaments \cite[e.g.][and reference therein; see also Pogosyan et al, in prep. for a  study of the connectivity of the cosmic web]{aubertetal04}. The key role played by this anisotropic environment in galaxy formation is increasingly underlined. For instance, it has been observed that the properties of galaxies --morphology, colours, luminosities, spins among others -- are correlated to their large-scale environment \citep[see][among many others]{oemler74,guzzo97,tempel&libeskind13,kovac14}.

Numerical simulations allow us to study the details of this large-scale structure of the Universe together with its impact on the formation and evolution of galaxies. Using N-body simulations, \cite{hahnetal07,gayetal10,metuki15} found that the properties of dark matter halos such as their morphology, luminosity, colour and spin parameter depend on their environment as traced by the local density, velocity and tidal field. In addition to scalar quantities, it also appears that their shape and spin are correlated to the directions of the surrounding filaments and walls both in dark matter \citep[see for instance][]{aubertetal04,bailin&steinmetz05,brunino07,calvoetal07,sousbie08,pazetal08,codisetal12,aragon14} and hydrodynamical simulations \citep{navarro04,hahn10,dubois14}. 

Analytical works provide important insights to understand the results of those simulations in the quasi-linear regime. As already pointed out, the theory of constrained random fields is an important tool that allows analytical calculations in the linear or weakly non-linear regime which is effective at large scales or early times in the Universe.
Virialised halos are the highly non-linear result of gravitational dynamics. They tend to form in the high-density peaks of the density field by gravitational instability and as such represent a biased tracer of the density field \citep{kaiser84,BBKS}. Peak statistics has focused a lot of attention in the recent years as it provides a unique way to analytically study the statistics of halos from their spatial distribution to their mass function \citep{2012MNRAS.426.2789P} or their spin \citep{ATTT}, at least for rare enough objects \citep{2011MNRAS.413.1961L}.

Despite clear evidence from numerical simulations, the detection of filaments and cold flows is still a debated but crucial issue as filamentary flows are often depicted as the solution to the missing baryons problem \citep{1992MNRAS.258P..14P,1998ApJ...503..518F,2001ApJ...552..473D,2012ApJ...759...23S}. In particular, gravitational lensing has emerged as a potential powerful probe of the filamentary cosmic web despite being challenging because of the systematics and the weakness of the signal \citep{2005A&A...440..453D,2010MNRAS.401.2257M,2016A&A...590A..69M}. 

{ {Gravitational lensing is related to the projected density integrated along the line of sight from distant source to the observer. The so-called convergence $\kappa$ is proportional to the projection of the density contrast $\delta$, and, as such, it inherits its statistical properties. In particular, projection will tend to wash  the non gaussianities of the $\delta$ field out. One would therefore try and enhance the importance of the filamentary structure by looking at the statistical properties of the convergence field at the vicinity of the rarest, most singular, events, which are the clusters at the nodes of the web. }}
In this work, we quantify the amount of symmetry of the matter distribution around clusters of galaxies by means of the aperture multipolar moments { { of the convergence field}} \citep{SB97} and their power spectrum. In particular, this tool should allow us to detect the signature of filaments feeding galaxy clusters in weak lensing surveys. This paper aims to do the theory of this observable in the Gaussian regime while a companion paper  {\citep{gouin}} explores the fully non-linear regime by analyzing clusters of galaxies within cosmological N-Body simulations.

This works complements in two dimensions the 3D harmonic analysis of infall at the Virial radius presented in \cite{aubertpichon2007}. The paper is organized as follows.
Section~\ref{sec:formalism} describes the mathematical formalism from the general definition of multipolar moments to the statistical description of peaks in Gaussian random fields (GRF hereafter) and their impact on the statistics of the multipolar moments.
Section~\ref{sec:GRF} then compares the predictions to measurements in Gaussian random fields.
Section~\ref{sec:sub} studies the effect of substructures and 
Section~\ref{sec:wr} adds a generic radial weight function.
 {We describe the weakly non-linear evolution of the multipolar moment in Sect.~\ref{sec:WNL}.}
Finally, we give preliminary conclusions of this work in Sect.~\ref{sec:conclusion} and propose possible follow-up developments.
A statistical characterisation of the geometry of peaks for 2D Gaussian random fields is given in App.~\ref{app:geom}.

\section{Formalism}
\label{sec:formalism}
\subsection{Aperture multipolar moments}
{ {The focus of this paper lies in the azimuthal mass distribution at various scales
around massive galaxy clusters.
 For a thin gravitational lens plane,
the convergence $\kappa$  at a given
position $\vec{r}$ in the sky corresponds to the projected excess surface density expressed
in units of the so-called critical density $\Sigma_{\rm crit}$
\begin{equation}
\kappa(\vec{r}) = \frac{1}{\Sigma_{\rm crit}} \int \der z
\left( \rho(\vec{r},z) - \overline{\rho} \right) \,,
\end{equation}
with the convention that the line-of-sight corresponds to the $z$-axis
and the plane of the sky $\vec{r}$ vector can be defined by polar
coordinates $(r,\varphi)$. The critical density involves distance
ratios between a fiducial source at an angular diameter distance
$D_{\rm s}$, the distance to the lensing mass $D_{\rm l}$ and the
distance between the lens and the source $D_{\rm ls}$
\begin{equation}\label{eq:scrit}
  \Sigma_{\rm crit} = \frac{c^2}{4 \pi G} \frac{ D_{\rm s}}{ D_{\rm l}
    D_{\rm ls}}\,.
\end{equation}
On cosmological scales, the thin lens approximation is generally not valid and the integrated deflections experienced by light rays as they travel from the source to the observer requires numerical treatment but for most cosmological applications the integration of the deflections along the unperturbed light rays \citep[so-called Born approximation, see eg][]{BS01} yields a linear integral relation between the convergence $\kappa$ and the density contrast $\delta$. For a known time-varying\footnote{where time variation is captured by an explicit dependence on comoving distance $\chi$} three-dimensional power spectrum $P_\delta(\vec{k},\chi)$, and for a given source plane redshift $z_{\rm s}$, one can thus write the convergence power spectrum $P_\kappa(\vec{\ell},z_{\rm s})$ by means of the Limber approximation \citep{blandford91,miralda91,K92,BS01,Simon+07}
\begin{equation}
P_\kappa(\ell,z_s) = \frac{9}{4} \Omega_m^2 \left( \frac{H_0}{c} \right)^4 \int_0^{\chi_{s}} \der \chi \frac{(\chi_s - \chi)^2}{\chi_s^2 } \frac{P_{\delta}\left({\ell}/{\chi},\chi \right)}{a^2(\chi)} \;.
\label{eq:pk_ell}
\end{equation}
}}
Following early works by \cite{SB97}, we define the aperture multipolar moments of the convergence (projected surface mass density) field $\kappa$ as
\begin{equation}
 Q_m = \int_0^\infty \der r\,  r^{1+m} w_m(r) \int_0^{2\pi} \der
  \varphi\,  {\rm e}^{i m \varphi} \kappa(r,\varphi)\,,
\end{equation}
with a radial weight function $w_{m}(r)$ commonly defined on a compact support. Those multipoles aim to quantify possible asymmetries in the mass distribution as probed by gravitational lensing. 

\begin{figure*}
\centering
\includegraphics[width=0.96\columnwidth]{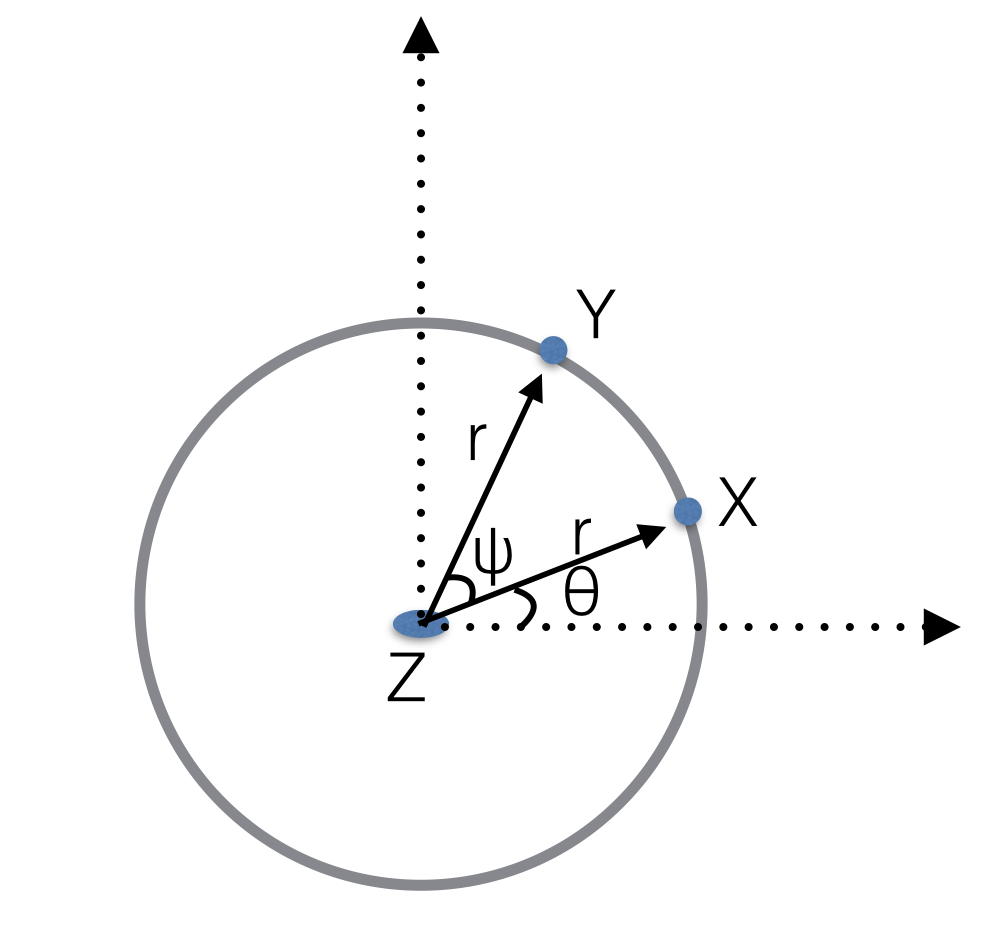} \hskip 1cm
\includegraphics[width=0.84\columnwidth]{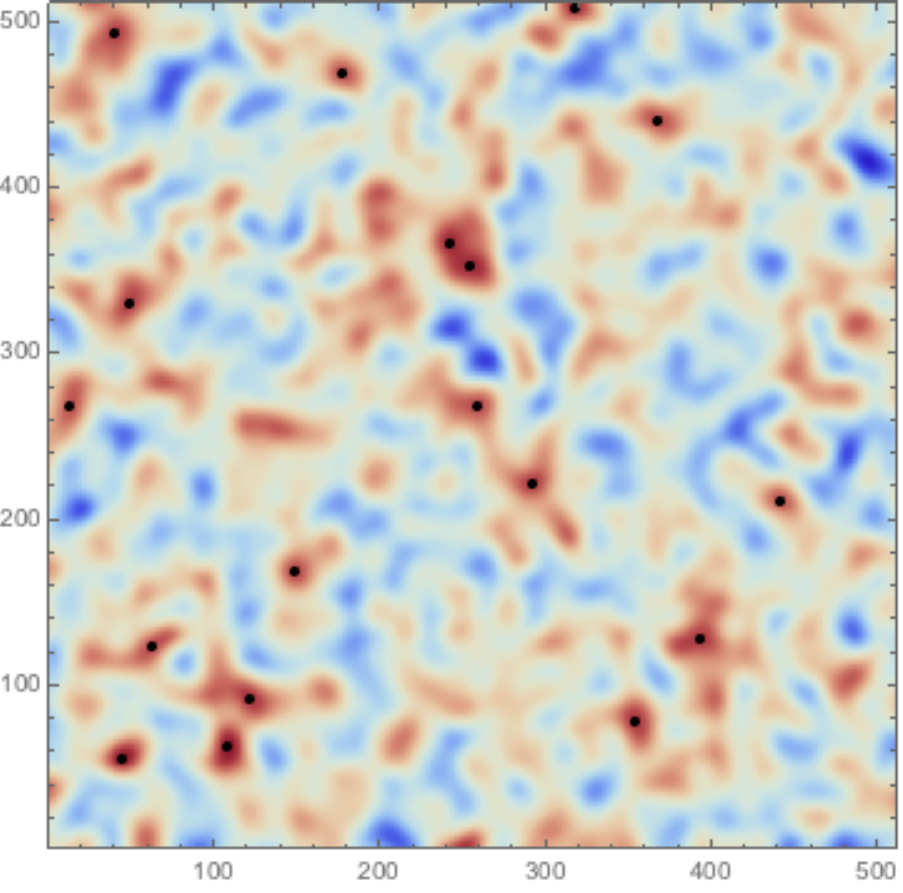} 
\caption{Left-hand panel: This paper aims at describing the angular distribution of a 2D Gaussian field $\kappa$ around a peak in ${\bf r}_{z}$. We will therefore consider two points on the annulus at a distance $r$ from the central peak. Their respective cartesian coordinates are ${\bf r}_{x}=r(\cos \theta,\sin\theta)$ and ${\bf r}_{y}=r(\cos \theta+\psi,\sin\theta+\psi)$. In particular, we will compute the expectation value of the product of the field in those two locations on the annulus given a central peak. Right-hand panel : example of such a 2D Gaussian random field with  {a power-law power spectrum with spectral index $n_{s}=0$}. Peaks of height $\nu=3\pm0.5$ are highlighted with black dots. We hereby investigate the polar distribution of the field around such peaks.\label{fig:frame}}
\end{figure*}

The covariance between multipolar moments can straightforwardly be written as
\begin{equation}\label{eq:mpolecov}
  \langle Q_n Q_m^* \rangle  = 2 \pi \, i^{n-m} \int k\der k\, U_n(k)U_m(k)\, P(k) \,.
\end{equation}
where $U_n(\ell)$ is the Hankel transform of the radial weight function
\begin{equation}\label{eq:weightHankel}
   U_m(\ell) = \int r \der r\, r^m w_m(r)  J_m( \ell r)\,,
\end{equation}
$J_m(x)$ are the first kind Bessel functions
and $P(k)$ is the power spectrum of the two-dimensional random field $\kappa$.

In a suite of papers (including  {\cite{gouin}} and Gavazzi et al, in prep.), we propose to use the full statistics of these multipolar moments around clusters of galaxies.
The covariance of the aperture multipolar moments in specific locations of space, such as the vicinity of clusters, becomes
\begin{multline}
\label{eq:def-multipoles}
 \langle Q_n Q_m^* |{\rm clusters}\rangle  =
 \iint\limits_0^\infty r\der r\, r'\der r' \,  \iint\limits_0^{2\pi} \der
  \varphi\der  \varphi'\, \,  r^{n} w_n(r)  r'^{m} w_m(r') \\
 \times   {\rm e}^{i (n \varphi - m\varphi')}\langle
  \kappa(r,\varphi) \kappa(r',\varphi') |{\rm clusters} \rangle\,,
  \end{multline}
where $ \langle
  \kappa(r,\varphi) \kappa(r',\varphi') | {\rm clusters}  \rangle$ is a constrained two-point correlation function as we impose a cluster at the origin of the polar coordinate system.
  
  In order to develop a physical intuition of the effect of this cluster constraint on the statistics of the multipolar moments, we propose in this paper to study analytically this observable for a Gaussian random field in which clusters are identified as high peaks.
 To simplify the problem, we drop the radial weight function and focus on Gaussian random fields smoothed with a Gaussian kernel on a given scale $R$.
 In what follows, we will investigate the angular distribution of a Gaussian random field around a peak. We therefore need to study the joint statistics of the field in three locations of space (the location of the peak and two arbitrary points on the circle at a distance $r$ away from the central peak). In addition, according to the peak theory originally developed in \cite{BBKS}, we need to consider the field, its first and second derivatives at the location of the peak. In Sect.~\ref{sec:JPDF}, we will first present the result for the joint PDF of those random variables before computing the resulting multipolar decomposition around a central peak in Sect.~\ref{sec:multipoles}. 

\subsection{Three-point statistics of the field and its  derivatives}
\label{sec:JPDF}

\begin{figure*}
\centering
\includegraphics[width=0.98\columnwidth]{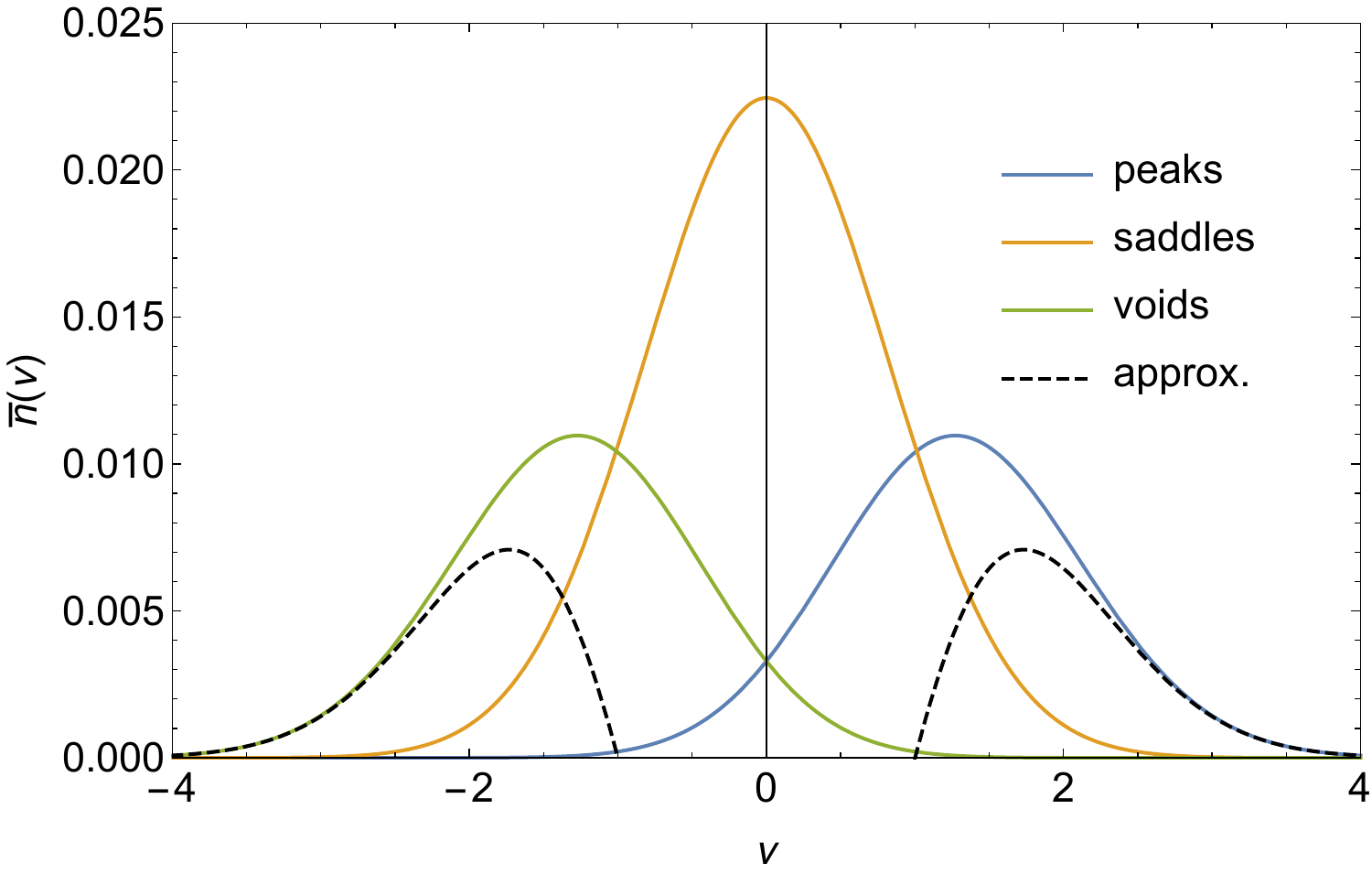} \hskip 0.5cm
\includegraphics[width=0.98\columnwidth]{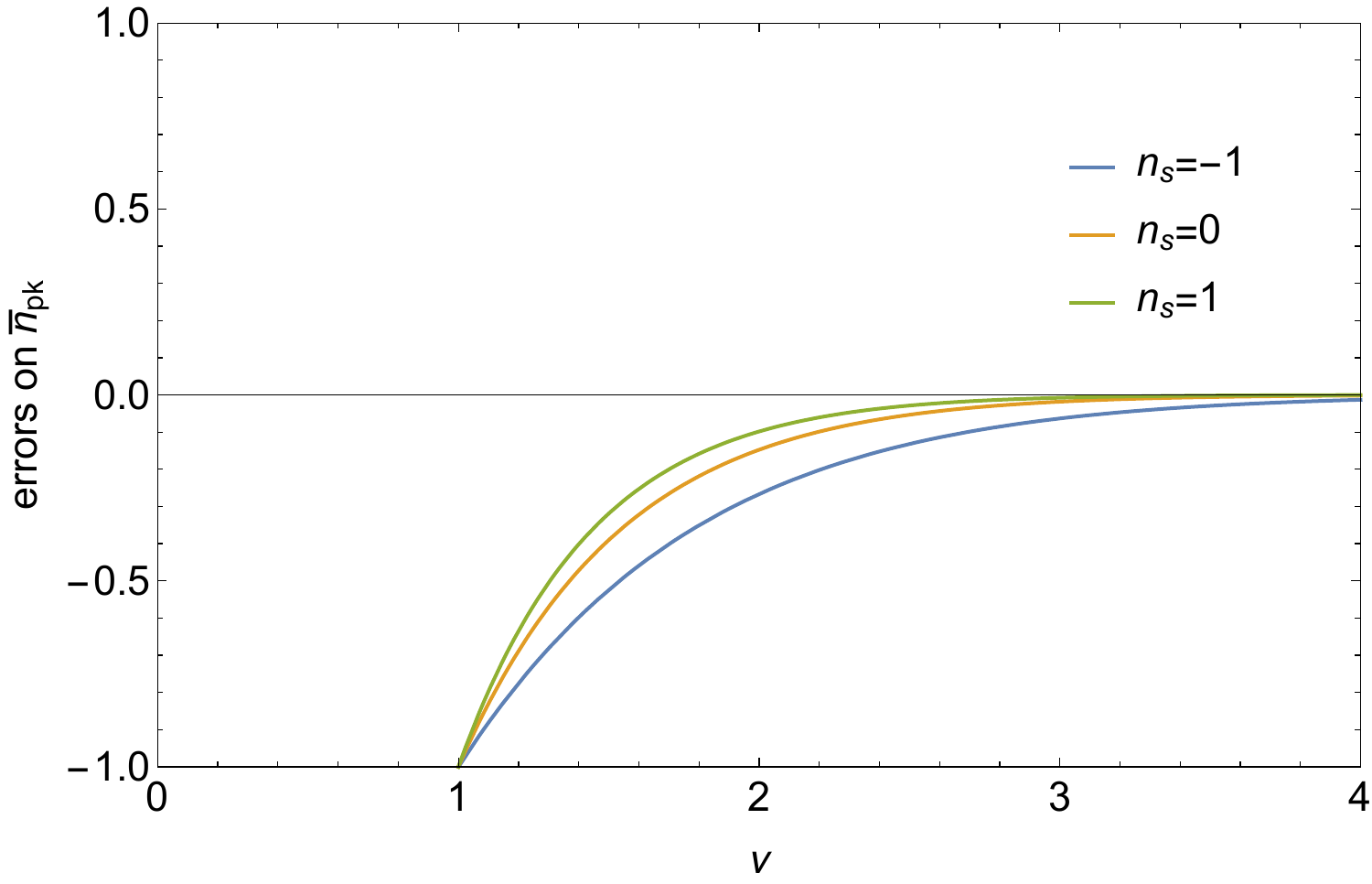}
\caption{Left-hand panel: Number density of minima, saddle points and peaks compared to the approximation of rare events in units of $R_{\star}^{-2}$. Right-hand panel : relative error on the number density of peaks of height $\nu$ when using the rare event approximation instead of the exact result. Different colours correspond to different spectral indices as labeled.
\label{fig:npk}}
\end{figure*}

For a given two-dimensional Gaussian field $\kappa$ (for example, the projected density contrast), we define 
the moments
\begin{align}
{\sigma_0}^2 &= \langle \kappa^2 \rangle, 
& {\sigma_1}^2 &= \langle \left( \nabla \kappa \right)^2 \rangle, 
& {\sigma_2}^2 &= \langle (\Delta \kappa)^2 \rangle.
\end{align}
From these moments, we will use two characteristic lengths
$R_0 ={\sigma_0}/{\sigma_1}$ and $R_\star = {\sigma_1}/{\sigma_2}$, as well as the 
spectral parameter
\begin{equation} 
\gamma=\frac{{\sigma_1}^2}{\sigma_0 \sigma_2}.
\end{equation}
Let us now define the following normalised random variables
\begin{align} 
x&=\frac{1}{\sigma_0} \kappa, 
& x_i&=\frac{1}{\sigma_1} \nabla_i \kappa, 
& x_{ij} &=\frac{1}{\sigma_2} \nabla_i \nabla_j \kappa,
\end{align}
which have unit variance by construction.

In what follows, ${\cal P}(\mathbf{X})$ denotes the one-point probability density (PDF) and
${\cal P}(\mathbf{X},\mathbf{Y},\mathbf{Z})$ designates the joint PDF for the normalized 
field and its derivatives, $\mathbf{X}=\{x\}$, $\mathbf{Y}=\{y\}$ and $\mathbf{Z}=\{z,z_{i},z_{ij}\}$,
at three prescribed comoving locations (${\bf r}_{x}$,${\bf r}_{y}$ and ${\bf r}_{z}$) separated by a distance
$r=|{\bf r}_{x}-{\bf r}_{z}|=|{\bf r}_{y}-{\bf r}_{z}|$ so that we are considering the density field in two locations, ${\bf r}_{x}$ and ${\bf r}_{y}$ on a same infinitely thin shell around the central peak in ${\bf r}_{z}$ -- see also the left-hand panel of Fig.~\ref{fig:frame} --. The right-hand panel of Fig.~\ref{fig:frame} shows a Gaussian random field and the position of its peaks. This paper aims to investigate the angular matter distribution around those peaks.

For a Gaussian field (in particular cosmic fields at early times or large scales), the
joint PDF is a multivariate normal distribution
\begin{equation}
{\cal N}(\mathbf{X},\mathbf{Y},\mathbf{Z})= \frac{\exp
\left[-\frac{1}{2}
\left(\begin{array}{c}
 \mathbf{X} 
\\
\mathbf{Y} 
 \\
\mathbf{Z} 
 \\
 \end{array} \right)^{\rm T}
 \cdot
  \mathbf{C}
 ^{-1}\cdot \left(\begin{array}{c}
 \mathbf{X} 
\\
\mathbf{Y} 
 \\
 \mathbf{Z} 
 \\
\end{array} \right) \right]
}
{{\rm det}|\mathbf{C}|^{1/2} \left(2\pi\right)^{\rm (6+3d+d^{2})/4 }} \,, 
\label{eq:defPDF}
\end{equation} 
where $d$ is the dimension -- $d=2$ here -- and $\mathbf{C}$ is the covariance matrix which depends on the separation vectors only because of homogeneity
\begin{equation}
\mathbf{C}=\left(\begin{array}{ccc}
1&\left\langle xy \right\rangle &\mathbf{C}_{\mathbf {XZ}}
\\
\left\langle xy \right\rangle &1&\mathbf{C}_{\mathbf{ YZ}}
 \\
 \mathbf{C}_{\mathbf {XZ}}^{\rm T} &\mathbf{C}_{\mathbf {YZ}}^{\rm T} &\mathbf{C}_{\mathbf{ ZZ}}\\
\end{array}
\right),
\end{equation}
with
\begin{eqnarray}
&&\mathbf{C}_{\mathbf {XZ}}=(\left\langle xz \right\rangle,\left\langle xz_{1} \right\rangle,\left\langle xz_{2} \right\rangle,\left\langle xz_{11} \right\rangle,\left\langle xz_{12} \right\rangle,\left\langle xz_{22} \right\rangle),\\
&&\mathbf{C}_{\mathbf {YZ}}=(\left\langle yz \right\rangle,\left\langle yz_{1} \right\rangle,\left\langle yz_{2} \right\rangle,\left\langle yz_{11} \right\rangle,\left\langle yz_{12} \right\rangle,\left\langle yz_{22} \right\rangle),\\
&&\mathbf{C}_{\mathbf {ZZ}}=\left(
\begin{array}{cccccc}
 1 & 0&0&-\gamma/2&0&-\gamma/2   \\
0 & 1/2 & 0&0&0&0 \\
0 & 0&1/2 &0&0&0 \\
-\gamma/2&0&0&3/8&0&1/8\\
0&0&0&0&1/8&0\\
-\gamma/2&0&0&1/8&0&3/8
\end{array}
\right).
\end{eqnarray}

For instance, for a 2D power-law power spectrum with spectral index $n_{s}$ smoothed with a Gaussian filter ($r$ is now the separation in units of the Gaussian smoothing length)
\begin{eqnarray}
&&\left\langle xz \right\rangle=\, _1F_1\left(\frac{n_{s}}{2}+1;1;-\frac{r^2}{4}\right)\equiv \xi(r),\\
&&\left\langle xy \right\rangle=\xi(|{\bf r}_{x}-{\bf r}_{y}|=2 r \sin( \psi/2)),\\
   &&\left\langle x\nabla z \right\rangle=\frac{\sqrt{n_{s}+2}}{2\sqrt 2} \, _1F_1\left(\frac{n_{s}}{2}+2;2;-\frac{r^2}{4}\right)\mathbf{r},
     \end{eqnarray}
     \begin{eqnarray}
&&\left\langle xz_{11} \right\rangle=-\frac{\gamma}{2} \left[2 \cos^2(\theta +\psi ) \, _1F_1\left(\frac{n_{s}}{2}+2;1;-\frac{r^2}{4}\right)\right.\\
&& \left.\hskip 1.8cm-\cos (2 (\theta +\psi )) \,
   _1F_1\left(\frac{n_{s}}{2}+2;2;-\frac{r^2}{4}\right)\right],\\
   &&\left\langle xz_{12} \right\rangle=-\frac{r^{2}\gamma(n_{s}+4)}{32}  \sin(2(\theta +\psi )) \, _1F_1\left(\frac{n_{s}}{2}+3;3;-\frac{r^2}{4}\right),\\
   &&\left\langle xz_{22} \right\rangle=-\frac{\gamma}{2} \left[2 \sin ^2(\theta +\psi ) \, _1F_1\left(\frac{n_{s}}{2}+2;1;-\frac{r^2}{4}\right)\right.\\
&& \left.\hskip 1.8cm+\cos (2 (\theta +\psi )) \,
   _1F_1\left(\frac{n_{s}}{2}+2;2;-\frac{r^2}{4}\right)\right].
\end{eqnarray}
Here $_1F_1(a;b;z)$ is the confluent hypergeometric function, $\xi$ is the two-point correlation function of the field and the spectral parameter reads $\gamma=\sqrt{(n_{s}+2)/(n_{s}+4)}$.
The correlation matrix $\mathbf{C}_{\mathbf {YZ}}$ is obviously the same as $\mathbf{C}_{\mathbf {XZ}}$ once $\psi$ has been set to zero.

\subsection{The central peak condition}
Eq.~(\ref{eq:defPDF}) is sufficient to compute the expectation of any quantity involving the 
fields and its derivatives up to second order in three different locations. This is the case if one wants to implement a peak condition at the ${\bf r}_{z}$ location. Indeed, following \cite{Longuet,Adler81,BBKS}, this peak constraint reads $|\det z_{ij}|\delta_{\textrm D}(z_{i})\Theta_{H}(-\lambda_{i})$ where $\delta_{D}(z_{i})\equiv\delta_{D}(z_{1})\delta_{D}(z_{2})$ is a product of Dirac delta functions which imposes the gradient to be zero, $\Theta_{H}(-\lambda_{i})\equiv\Theta_{H}(-\lambda_{1})\Theta_{H}(-\lambda_{2})$ an Heaviside function forcing the curvatures (equivalently the eigenvalues of the Hessian matrix $\lambda_{i}$) to be negative. The factor $|\det z_{ij}|=|z_{11}z_{22}-z_{12}^{2}|$ encodes the volume associated to each peak, in other words the Jacobian which allows us to go from a smoothed field distribution to the discrete distribution of peaks. The rareness of the peak $\nu$ can also be imposed by adding a factor $\delta_{\textrm D}(z-\nu)$. We will therefore denote $n_{\rm pk}(\mathbf Z)$ the localized density of peaks
\begin{equation}
\label{eq:constraintpk}
n_{\rm pk}(\mathbf Z)=\frac{1}{R_{\star}^{2}}|\det z_{ij}|\delta_{\textrm D}(z_{i})\Theta_{H}(-\lambda_{i})\delta_{\textrm D}(z-\nu)\,.
\end{equation}

\begin{figure*}
\centering
\includegraphics[width=0.65\columnwidth]{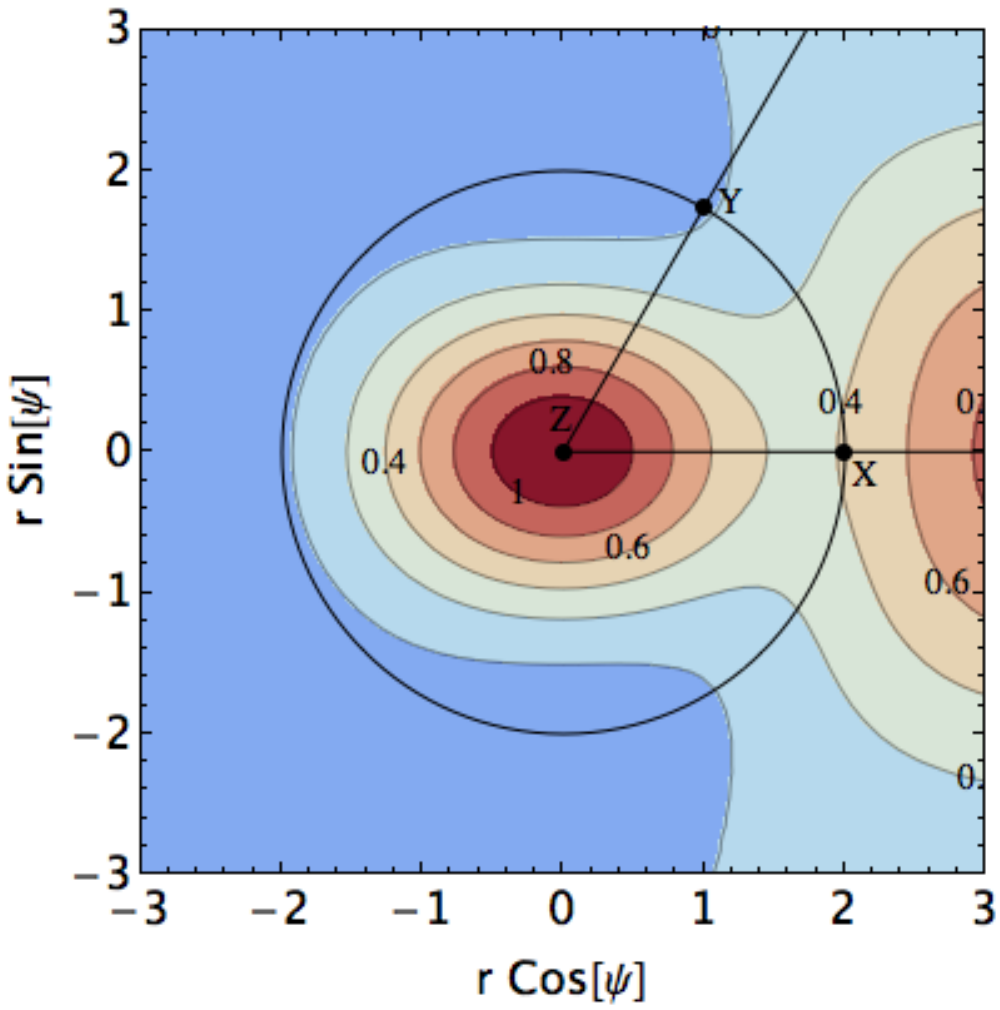} \;\;
\includegraphics[width=0.65\columnwidth]{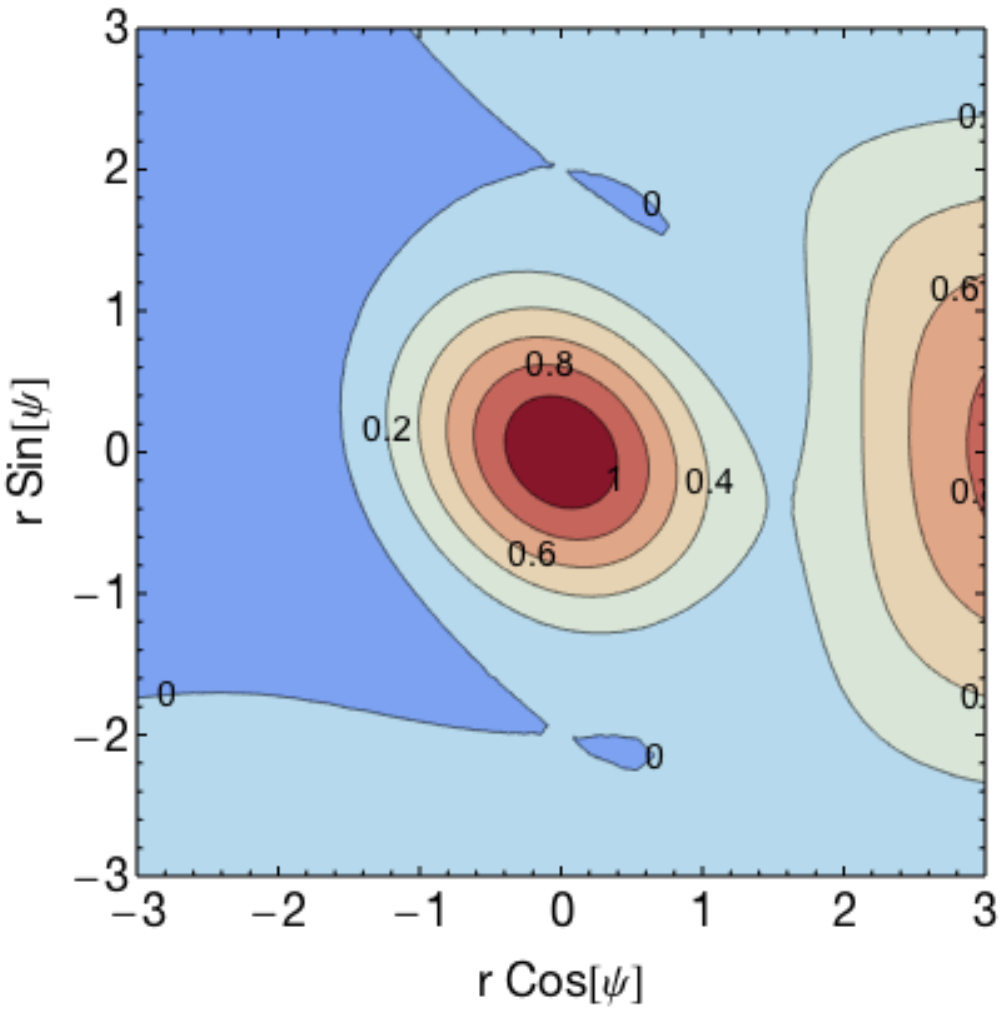} \;\;
\includegraphics[width=0.65\columnwidth]{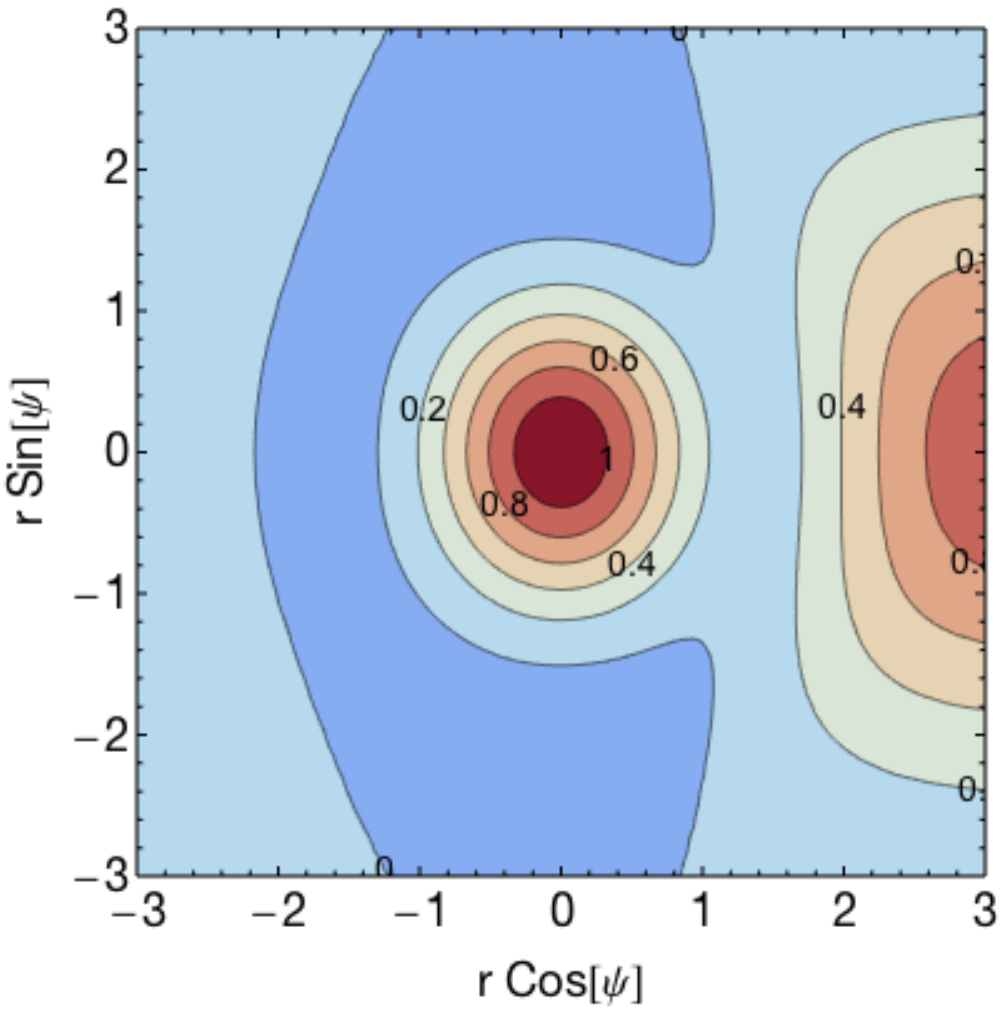} \\
\includegraphics[width=0.65\columnwidth]{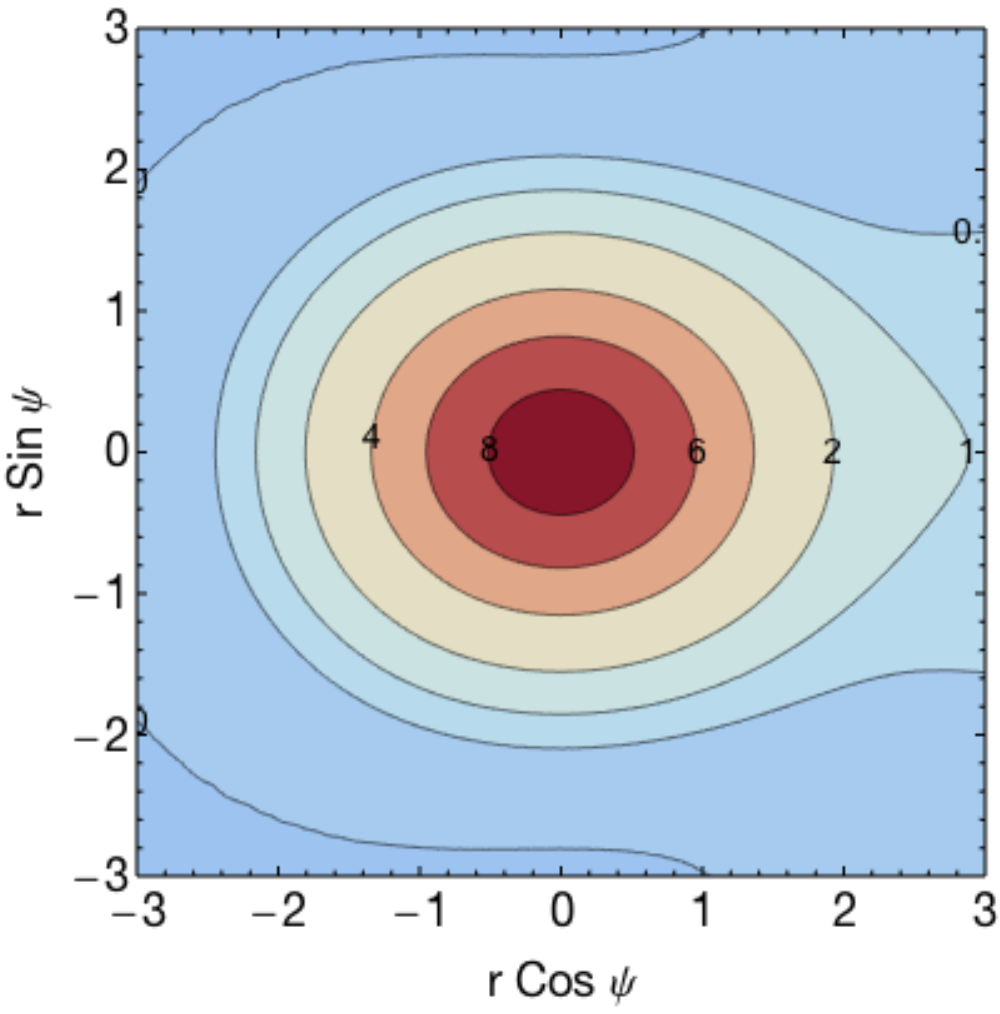} \;\;
\includegraphics[width=0.65\columnwidth]{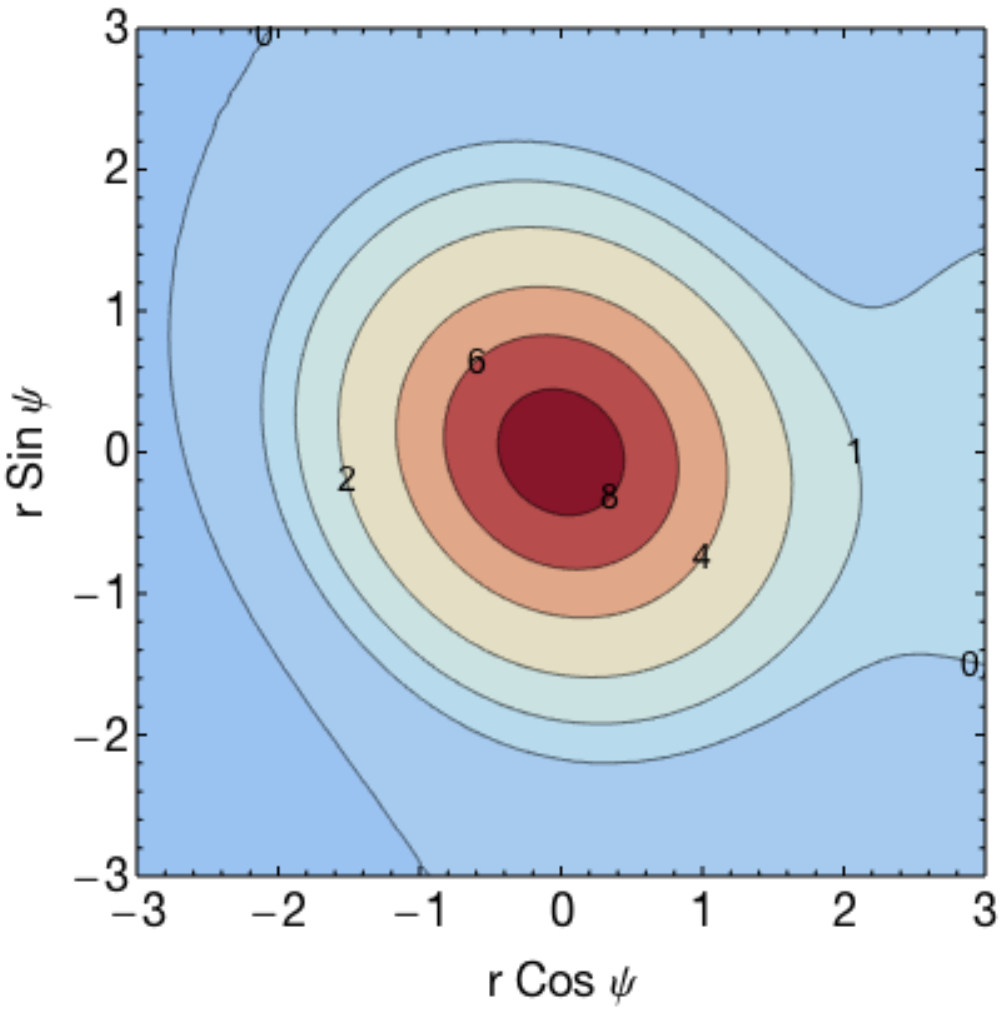} \;\;
\includegraphics[width=0.65\columnwidth]{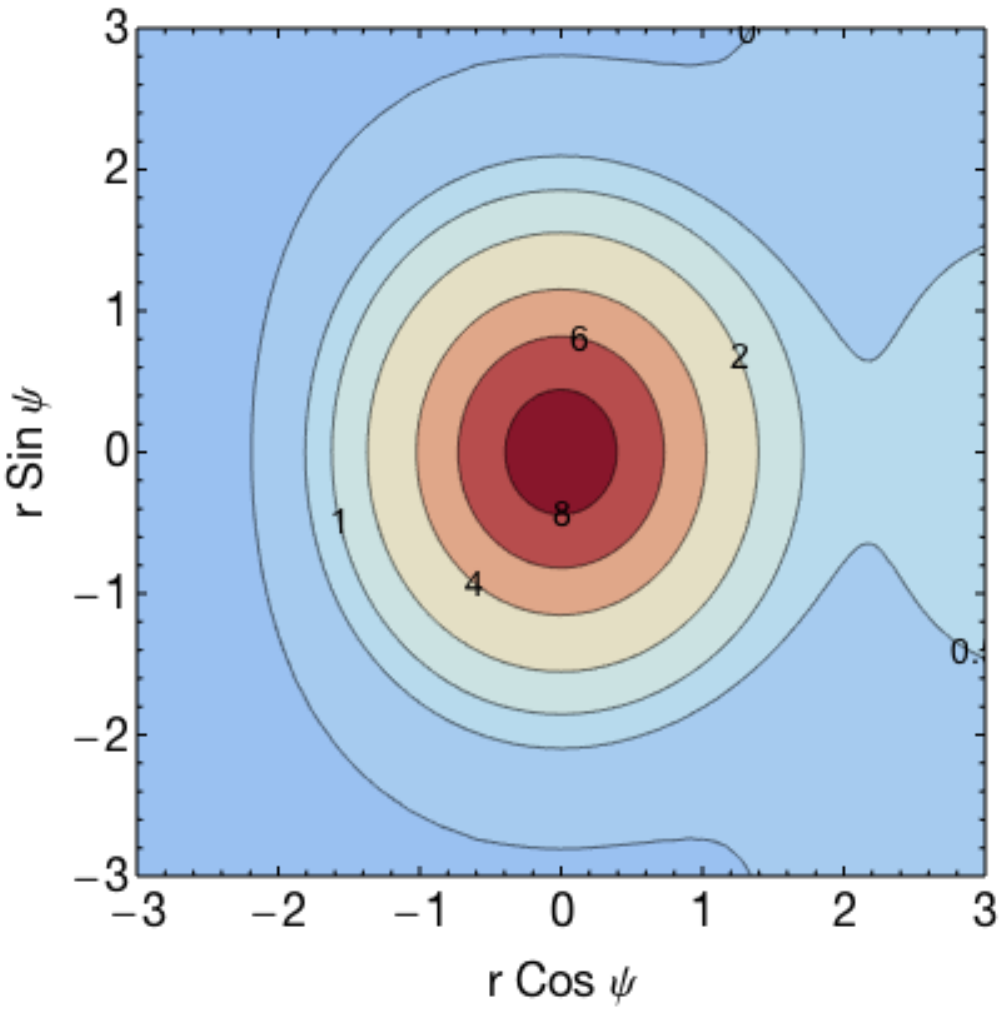} 
\caption{Top panels: expected two-point correlation function $\left\langle \kappa\kappa'|\textrm{pk}\right\rangle$ in units of $\sigma_{0}^{2}$ for a 2D power spectrum with spectral index $n_{s}=0$ and a central peak of height $\nu_{\star}=\sqrt {7/3}\gamma$ and eigenvalues $\lambda_{1\star}=(-\sqrt{7/3}+\sqrt{1/3})/2$ and  $\lambda_{2\star}=(-\sqrt{7/3}-\sqrt{1/3})/2$  {in ${\bf r}_z$}. Different values of $\theta$, the angle between the major axis of the ellipse (i.e smaller curvature) and the first point on the annulus, between 0 and $\pi/2$ are displayed from left to right. $\psi$ is the angle between ${\bf r}_{x}$ and ${\bf r}_{y}$  and $r$ is the separation to the central peak (in units of the smoothing length)  {so that the displayed value corresponds to the correlation function between this point, ${\bf r}_{y}$, and the one on the positive x-axis at the same radius, ${\bf r}_{x}$}. The values we chose here correspond to the most likely height and curvatures of a peak (and do not depend on the spectral index). Bottom panels: same as top panels for $\nu_{r}=3$. The corresponding most likely curvatures of the peak are  $\lambda_{1r}=-0.94$ and $\lambda_{2r}=-1.6$.
\label{fig}}
\end{figure*}

The most difficult part in the peak constraint is often to impose the sign of the curvatures and the positivity of the Jacobian which can prevent from getting analytical results as it is the case for 3D differential peak counts \citep{gay12} or peak-peak correlation functions (as described in \cite{Baldauf16} in one dimension and \cite{Regos1995} in three dimensions) which can only be solved numerically. A standard approximation to keep results  {analytical} is to drop this sign constraint and remove the absolute values of the determinant factor for high contrasts as one expects rare enough critical points to be essentially peaks. If this approximation is very accurate for one-point statistics, it may not be the case for ($N>1$)-point statistics. For instance, peak-peak correlation functions on small scales are not very well reproduced by this approximation even for large contrasts because the contribution from the other critical points actually dominates at small distance (there is at least one saddle point between two peaks!). However, in the context of this work, we impose the peak constraint in one location only and therefore the rare peak approximation is expected to be accurate for $\nu\gtrsim 2$.
  As an illustration, Fig.~\ref{fig:npk} displays the Gaussian mean number density of minima, saddle points and peaks (\cite{Longuet,Adler81} and later generalised to weakly non-Gaussian fields by \cite{pogo11})
 \begin{eqnarray}
 \bar n_{\rm pk/min}(\nu)&=&\frac{\gamma^{2}\sqrt 2\exp\left(-\frac{\nu^{2}}{2}\right)}{16\pi^{3/2}R_{\star}^{2}}(\nu^{2}-1)\!\left(\!1\pm{\rm erf}\!\left(\!\frac{\gamma\nu}{\sqrt{2(1-\gamma^{2})}}\!\right)\!\right)\!\nonumber\\
 &+&\!\!\!\!\frac{\sqrt 2\exp\left(-\frac{3\nu^{2}}{6-4\gamma^{2}}\right)}{16\sqrt{3-2\gamma^{2}}\pi^{3/2}R_{\star}^{2}}\!\left(\!1\pm{\rm erf}\!\left(\!\frac{\gamma\nu}{\sqrt{2(1-\gamma^{2})(3-2\gamma^{2})}}\!\right)\!\right)\!\nonumber\\
 &\pm&\frac{\sqrt{1-\gamma^{2}}}{8\pi^{2}}\gamma\nu\exp\left(-\frac{\nu^{2}}{2-2\gamma^{2}}\right)\nonumber\,,
 \\
  \bar n_{\rm sad}(\nu)&=&\frac{\sqrt 2\exp\left(-\frac{3\nu^{2}}{6-4\gamma^{2}}\right)}{8\sqrt{3-2\gamma^{2}}\pi^{3/2}R_{\star}^{2}}\nonumber\,,
 \end{eqnarray}
 and compares the latter to the high-$\nu$ approximation (related to the genus) $ \chi(\nu)=\langle\det z_{ij}\delta_{\textrm D}(z_{i})\delta_{\textrm D}(z-\nu)\rangle/R_{\star}^{2}$ which can be easily computed
 \begin{equation}
 \chi(\nu)=\frac{\gamma^{2}}{4\sqrt 2 \pi^{3/2}R_{\star}^{2}}\exp\left(-\frac{\nu^{2}}{2}\right)(\nu^{2}-1)\,.
 \end{equation}
 The relative error between the number density of peaks and its high-$\nu$ approximation is shown on the right-hand panel of Fig.~\ref{fig:npk}.

\subsection{Density correlations on the circle surrounding a central peak with given geometry}

\begin{figure}
\centering
\includegraphics[width=0.9\columnwidth]{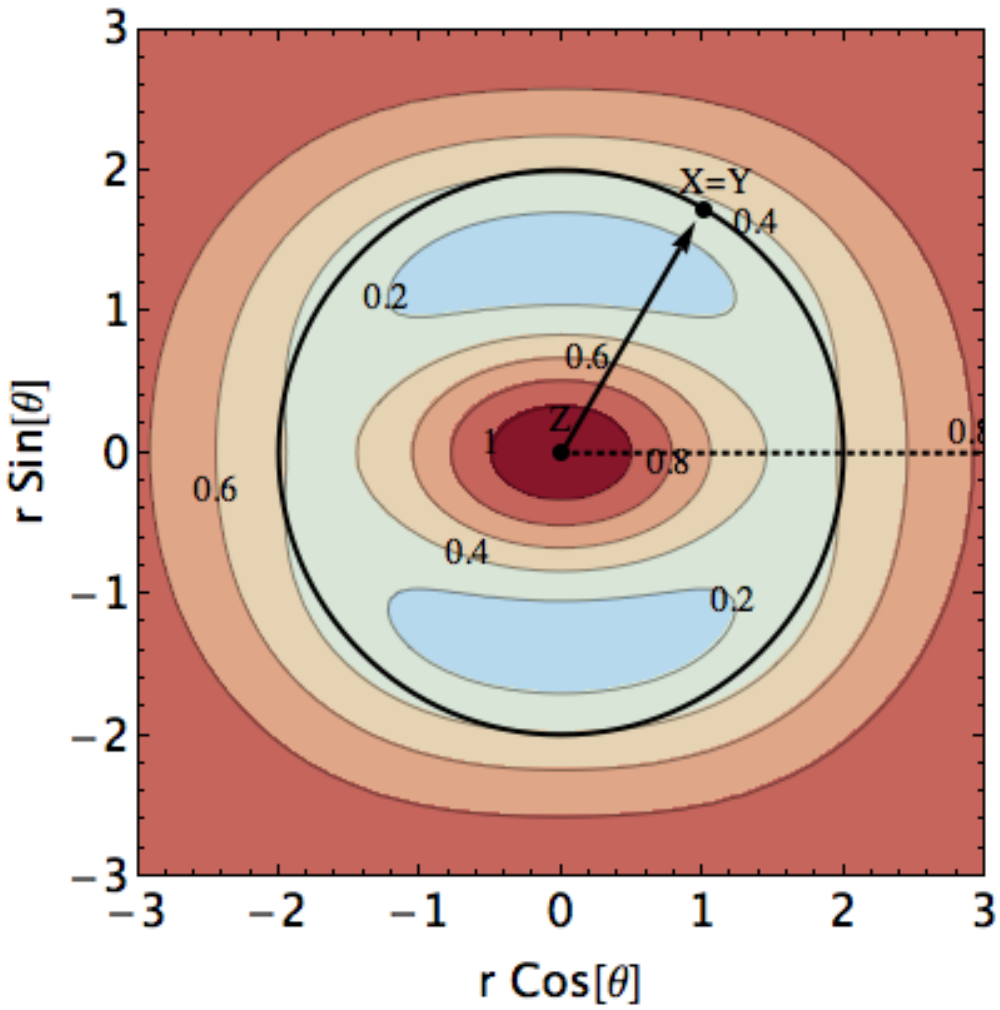}
\caption{ {Zero lag annulus correlation function $\left\langle \kappa(r,\theta)^2|{\rm pk}(\nu)\right \rangle$ in units of $\sigma_{0}^{2}$ for a central peak with height $\nu_{\star}=\sqrt {7/3}\gamma$ and eigenvalues $\lambda_{1\star}=(-\sqrt{7/3}+\sqrt{1/3})/2$ and  $\lambda_{2\star}=(-\sqrt{7/3}-\sqrt{1/3})/2$ in ${\bf r}_z$ for a Gaussian random field with power spectrum $P(k)\propto k^{0}$ smoothed with a Gaussian filter (similar to Fig.~\ref{fig} when the two points on the annulus are the same). The separation $r$ is given in units of the smoothing length. }
\label{fig:var}}
\end{figure}

The expected product of projected density $\kappa$ in two locations of space ${\bf r}_{x}$ and ${\bf r}_{y}$ such that ${\bf r}_{x}-{\bf r}_{z}=r(\cos\theta,\sin\theta)$ and ${\bf r}_{y}-{\bf r}_{z}=r(\cos(\theta+\psi),\sin(\theta+\psi))$  and given a
peak in ${\bf r}_{z}$ of height $\nu$ and curvatures $0>\lambda_{1}>\lambda_{2}$ along the first and second coordinates can be analytically computed.
For instance for a  {power-law} power spectrum with $n_{s}=0$ (and $\gamma=1/\sqrt{2}$), we get
\begin{multline}
\frac{\left\langle\kappa(r,\theta) \kappa(r,\theta+\psi)|\textrm{pk}\right\rangle}{\sigma_{0}^{2}}=
\xi(|{\bf r}_{x}-{\bf r}_{y}|)\\
+\exp\left(-\frac {r^{2}}{2}\right)\left[l_{0}+l_{2}r^{2}+l_{4}r^{4}\right],
\end{multline}
where
\begin{eqnarray}
l_{0}&=&(\nu^{2}-1),\\
l_{2}&=&[\nu^{2}+\sqrt 2\nu I_{1}(1-2 e \cos\psi\cos(2\theta+\psi))-\cos\psi ]/2,\\
l_{4}&=&\left[\nu^{2}-2\cos^{2}\psi+2\sqrt2 \nu I_{1}(1-2 e \cos\psi\cos(2\theta+\psi))\right.\nonumber\\
&&\left.+2I_{1}^{2}(1-2 e\cos(2\theta))(1-2 e\cos(2\theta+2\psi))\right]/16,
\end{eqnarray}
$I_{1}=\lambda_{1}+\lambda_{2}$ is the trace of the density Hessian at the location of the peak, $e=(\lambda_{2}-\lambda_{1})/(2I_{1})$ is the ellipticity of the peak {  and the unconstrained correlation function is
\begin{equation}
\xi(|{\bf r}_{x}-{\bf r}_{y}|)=\exp\left(-\frac {r^{2}}{2}(1-\cos\psi)\right).
\end{equation}
}
 {To start with, Fig.~\ref{fig:var} shows the zero lag contribution to the annulus correlation function (i.e when the two points are at the same location on the annulus, $\psi=0$) in the frame of the central peak.
As expected the amplitude of fluctuations around the peak have an ellipsoidal shape, more elongated along the smallest curvature $\lambda_{1\star}$. Note that this zero-lag annulus correlation is dominated by the square of the mean density profile at small separations and by the fluctuations at larger separations.}

Fig.~\ref{fig} then displays the full constrained correlation function on the annulus. We use different orientations of the pair (${\bf r}_{x},{\bf r}_{y}$) with regard to the axis of smaller curvature (corresponding to $\lambda_{1}$) of the central peak. The orientation of ${\bf r}_{x}$ is described by the angle $\theta$ which is taken to be 0, $\pi/4$ and $\pi/2$ from the left-hand  to the right-hand panel. On each plot, the angle between ${\bf r}_{x}$ and ${\bf r}_{y}$, namely $\psi$, vary between 0 and $2\pi$ and the separation to the central peak is described by the value $r$. We show the result for two different peak heights, the most likely value $\nu_{\star}=\gamma\sqrt{7/3}$ (top panels) and a rarer case $\nu_{r}=3$ (bottom panels) more relevant to our study. In each case respectively, we fix the peak curvatures to their most likely values $\lambda_{1\star}=(-\sqrt{7/3}+\sqrt{1/3})/2,\lambda_{2\star}=(-\sqrt{7/3}-\sqrt{1/3})/2$ and $\lambda_{1r}=-0.94,\lambda_{2r}=-1.6$ (we refer the reader to App.~\ref{app:geom} for a description of the most likely geometry of a peak). As expected, the product of density is larger when the separation vectors are close one to the other and aligned with the major axis of the peak. For the case of a rare peak (bottom panels), the prominence of the peak is obviously larger (increased magnitude and spatial extend of the peak). Conversely, the common peak, $\nu_{\star}$ occupies a smaller volume and is surrounded by two closer voids and peaks. In what follows, we do not fix the shape of the peak and therefore we marginalise over $\lambda_{1}$ and $\lambda_{2}$.

\subsection{Density correlations around a peak of specified height $\nu$}
\begin{figure}
\centering
\includegraphics[width=0.9\columnwidth]{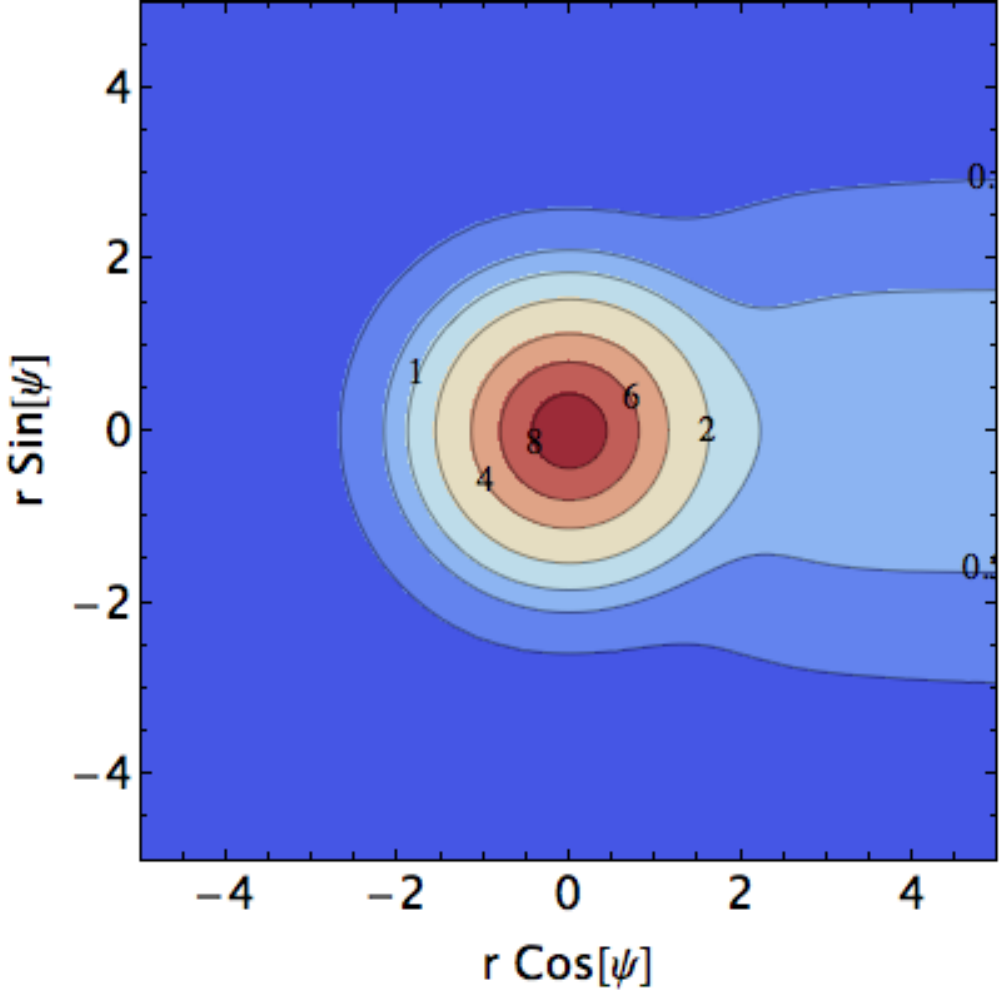}
\caption{ {Annulus correlation function $\left\langle \kappa(r,\theta) \kappa(r,\theta+\psi)|{\rm pk}(\nu)\right \rangle$ in units of $\sigma_{0}^{2}$ for a central peak with height $\nu=3$ in a Gaussian random field with power spectrum $P(k)\propto k^{0}$ smoothed with a Gaussian filter. The separation $r$ is given in units of the smoothing length. The angular anisotropy of the annulus correlation function will be quantified using a multipolar decomposition in Sect.~\ref{sec:multipoles}.}
\label{fig:kk-pk}}
\end{figure}

\begin{figure*}
\centering
\includegraphics[width=0.98\columnwidth]{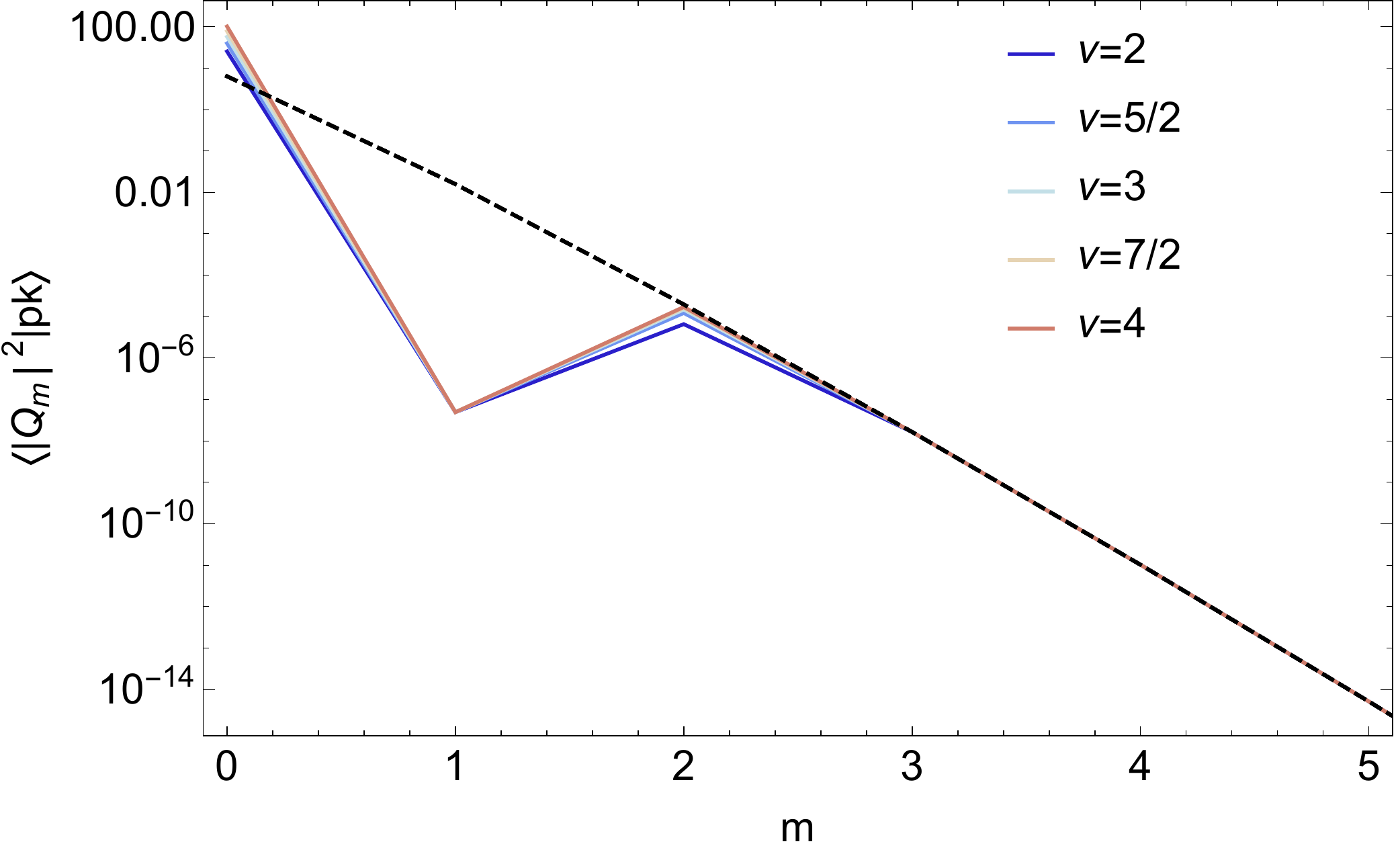} \hskip 0.5cm
\includegraphics[width=0.98\columnwidth]{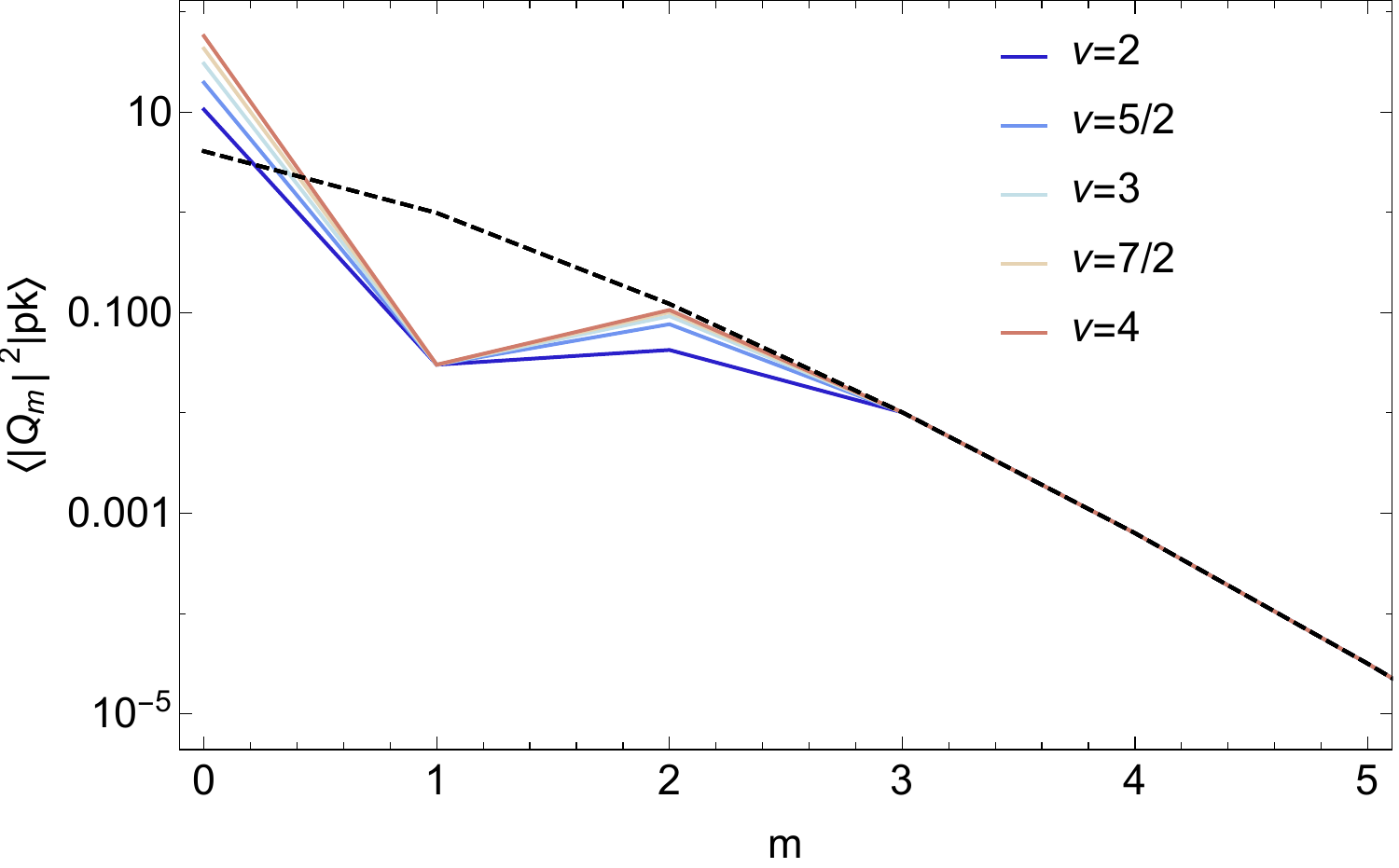} \\
\includegraphics[width=0.98\columnwidth]{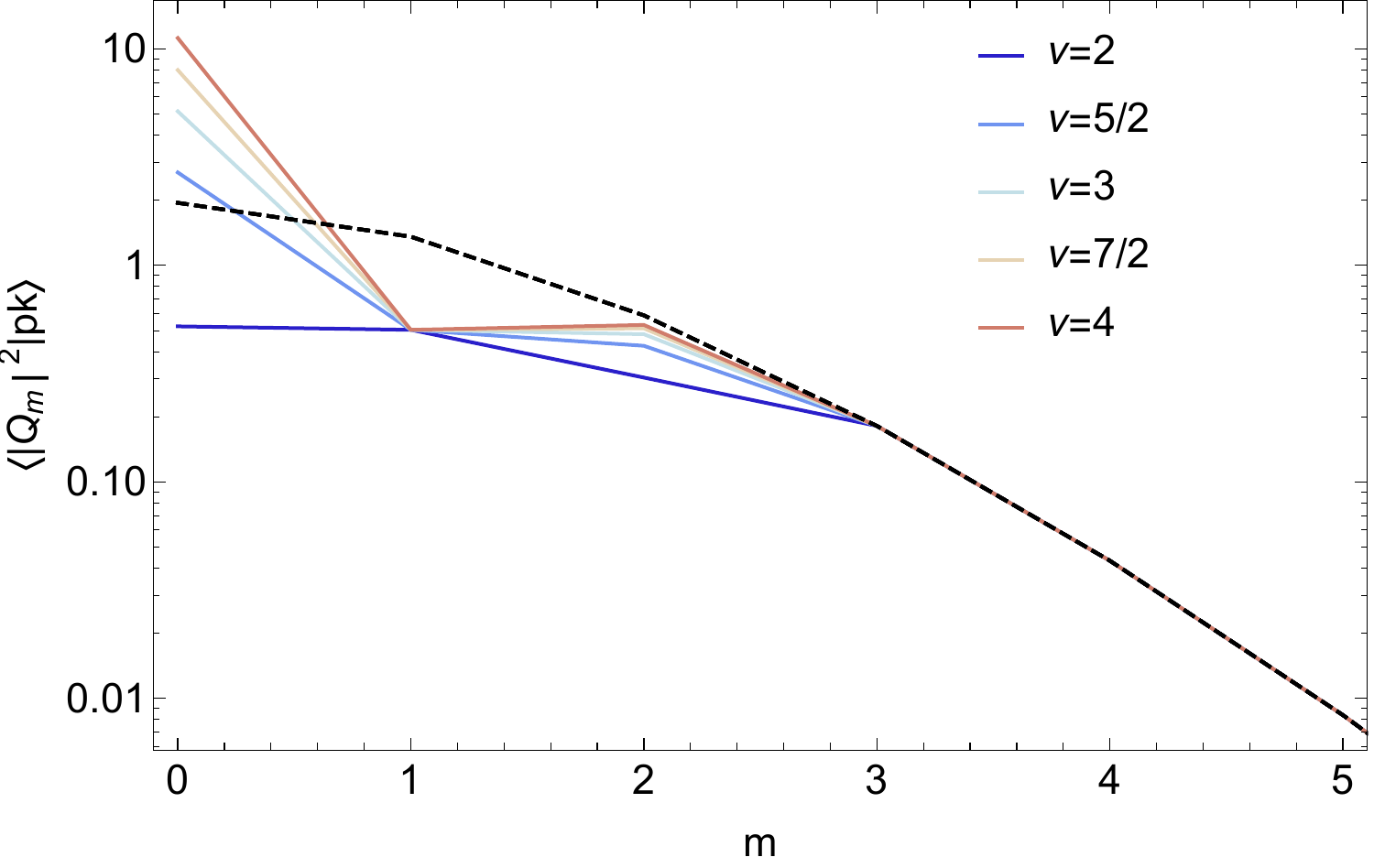} \hskip 0.5cm
\includegraphics[width=0.98\columnwidth]{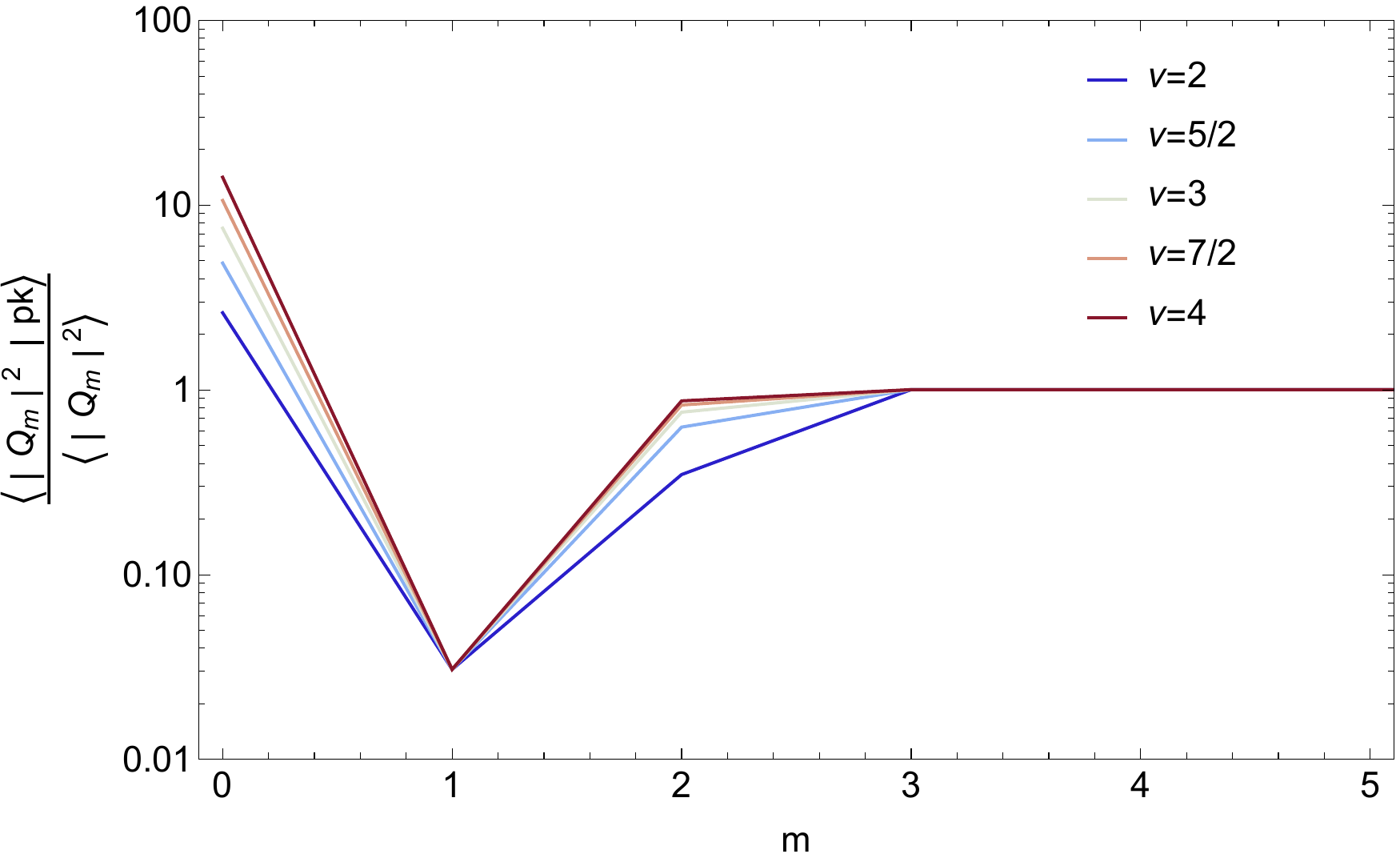} 
\caption{ {Multipoles $ |Q_{m}|^{2}|$  for a central peak with height $\nu=2$ to 4 as labeled in a Gaussian random field with power spectrum $P(k)\propto k^{0}$ smoothed with a Gaussian filter and on the annulus at a distance $r=0.1R$ (top left-hand panel), $R$ (top right), $2R$ (bottom left). The dashed line corresponds to the random case (where we do not impose a central peak). The bottom right-hand panel shows the ratio of those multipoles to the random case for $r=R$.
\label{fig:Qm}}}
\end{figure*}

If one wants to marginalise over the shape of the peak (which means integrating over the eigenvalues $\lambda_{1}$ and $\lambda_{2}$ in the range $\lambda_{2}<\lambda_{1}<0$), then the expected product of projected densities on the annulus (the annulus two-point correlation function) becomes
\begin{equation}
\frac{\left\langle \kappa(r,\theta) \kappa(r,\theta+\psi)|{\rm pk}(\nu)\right \rangle}{\sigma_{0}^{2}}=\frac{\left\langle x y \det(z_{ij}){\delta_{\textrm D}}(z-\nu){\delta_{\textrm D}}(z_{i}){\Theta_{H}}(-\lambda_{i}) \right\rangle}{\left\langle  \det(z_{ij}){\delta_{\textrm D}}(z_{i}){\Theta_{H}}(-\lambda_{i}) \right\rangle},\nonumber
\end{equation}
where we marginalize over all variables except $\nu$ which is fixed.
Unfortunately, this expression cannot be analytically computed. For sufficiently rare peaks (high $\nu$), we drop the constraint on the sign of the eigenvalues (high critical points are most of the time peaks) and an explicit expression for $\left\langle \kappa(r,\theta) \kappa(r,\theta+\psi)|{\rm pk}(\nu)\right \rangle$ can be obtained
\begin{multline}
\label{eq:kk-pk}
\frac{\left\langle \kappa(r,\theta) \kappa(r,\theta+\psi)|{\rm pk}(\nu)\right \rangle}{\sigma_{0}^{2}}\!=\!{  \xi(|{\bf r}_{x}-{\bf r}_{y}|)+\frac{2f_{21}^{2}}{\nu^{2}\!-\!1}+4f_{11}f_{21}}\\{  +\frac{\nu^{4}\!-\!6\nu^{2}\!+\!3}{\nu^{2}\!-\!1}f_{11}^{2}}\!
-\!\frac{n_{s}\!+\!2}{4}r^{2}f_{22}^{2}\cos\psi \!-\!\frac{2\cos(2\psi)}{\nu^{2}\!-\!1}(f_{21}\!-\!f_{11})^{2}\!,\!
\end{multline}
where $f_{ij}$ {  and the unconstrained correlation function $ \xi(|{\bf r}_{x}-{\bf r}_{y}|)$} are functions of the following Kummer confluent hypergeometric functions
\begin{eqnarray}
&&f_{ij}=\pFq{1}{1}{\frac {n_{s}} 2 +i}{j}{-\frac{r^{2}}{4}},\\
&&{  \xi(|{\bf r}_{x}-{\bf r}_{y}|)}=\pFq{1}{1}{\frac {n_{s}} 2 +i}{j}{-\frac{r^{2}}{2}(1-\cos\psi)}.
\end{eqnarray}

As an illustration, for a power spectrum $P(k)\propto k^{0}$, it becomes
\begin{multline}
\frac{\left\langle \kappa(r,\theta) \kappa(r,\theta+\psi)|{\rm pk}(\nu)\right \rangle}{\sigma_{0}^{2}}=\xi(|{\bf r}_{x}-{\bf r}_{y}|)+\exp\left(-\frac{r^2}{2}\right)\\
\times\frac{8 (\nu^{2}\!-\!1) ^2\!-\!8
   \nu ^2 r^2 \!+\!r^4\!-\!4 \left(\nu ^2\!-\!1\right) r^2 \cos \psi \!-\!r^{4}\cos 2 \psi }{8 \left(\nu ^2\!-\!1\right)},
\end{multline}
{  where $\xi(|{\bf r}_{x}-{\bf r}_{y}|)=\exp\left(-\frac {r^{2}}{2}(1-\cos\psi)\right)$ is the unconstrained correlation function on the annulus.}
The apparent singularity at $\nu=\pm1$ is due to our high $\nu$ approximation which breaks down in this regime as many $\nu=1$ critical points are not peaks but saddle points.
Fig.~\ref{fig:kk-pk} illustrates the behaviour of $\left\langle \kappa(r,\theta) \kappa(r,\theta+\psi)|{\rm pk}(\nu)\right \rangle$ for a central peak with height $\nu=3$.
 {Similarly to the case where the peak geometry is imposed, here the annulus correlation function is larger when the separation vectors are close one to the other and aligned with the major axis of the peak. The isocontours are close to spherical for small separations but become very anisotropic and elongated along the axis $\psi=0$ (when the two points overlap) at larger separations. }

\subsection{Multipoles around a peak of specified height $\nu$}
\label{sec:multipoles}

Once the two-point correlation function around a peak -- $\left\langle \kappa(r,\theta) \kappa(r,\theta+\psi)|{\rm pk}(\nu)\right \rangle$ -- is known, one can compute the corresponding multipolar moments that we define here as 
\begin{equation}
\left\langle |Q_{m}|^{2}|\textrm{pk}\right\rangle(r,\nu) \!=\!\int_{0}^{2\pi}\frac{\dd \psi}{2\pi\sigma_{0}^{2}} \left\langle\kappa(r,\theta) \kappa(r,\theta+\psi)|\textrm{pk}\right\rangle e^{\imath m \psi}.
\end{equation}
The result is again analytical. As expected, only the first three multipoles are modified by the peak condition,
the rest being unchanged
\begin{equation}
\left\langle |Q_{m}|^{2}|\textrm{pk}\right\rangle=\left\langle |Q_{m}|^{2}\right\rangle \textrm{for all } m\geq 3\,.
\end{equation}
For instance, for $P(k)\propto k^{0}$ power spectra, those multipoles read
\begin{eqnarray}
\left\langle |Q_{0}|^{2}|\,\textrm{pk}\right\rangle\!\!&=&\!\!\!\!\left\langle |Q_{0}|^{2}\right\rangle\!+\!\frac{r^{4}-8\nu^{2}r^{2}+8(\nu^{2}-1)^{2}}{8(\nu^{2}-1)}\exp\left(-\frac {r^{2}}{2}\right)\label{eq:Q0th},\\
\left\langle |Q_{1}|\,^{2}|\,\textrm{pk}\right\rangle\!\!&=&\!\!\left\langle |Q_{1}|^{2}\right\rangle-\frac 1 4 r^{2}\exp\left(-\frac {r^{2}}{2}\right),\\
\left\langle |Q_{2}|^{2}|\,\textrm{pk}\right\rangle\!\!&=&\!\!\left\langle |Q_{2}|^{2}\right\rangle-\frac 1 {16}\frac{r^{4}}{\nu^{2}-1}\exp\left(-\frac {r^{2}}{2}\right),\\
\left\langle |Q_{m}|^{2}|\,\textrm{pk}\right\rangle\!\!&\overset{ m\geq 3}{=}&\!\!\left\langle |Q_{m }|^{2}\right\rangle\equiv\exp\left(-\frac {r^{2}}{2}\right)I_{m}\left(\frac{r^{2}}{2}\right) \label{eq:Qmth},
\end{eqnarray}
where $I_{m}$ are the modified Bessel functions of the first kind. We note in particular that the correction to the monopole (resp. dipole, quadrupole) is maximal for $r=0$ (resp. $\sqrt 2$, 2).
It can easily be checked that the condition of zero gradient only affects the dipole, while the constraint on the peak height changes the monopole and the Hessian modifies both the monopole and quadrupole.

Fig.~\ref{fig:Qm} shows the amplitude of the multipoles for various peak heights and separations. There is a significant drop of power  in the dipole while the change in the monopole and quadrupole is much less pronounced. The dependance on the peak height is rather small. Those predictions will be checked against GRF realizations in Sect.~\ref{sec:GRF}.

\subsection{Dependence on the slope of the power spectrum}
\label{sec:ns}
\begin{figure}
\centering
\includegraphics[width=\columnwidth]{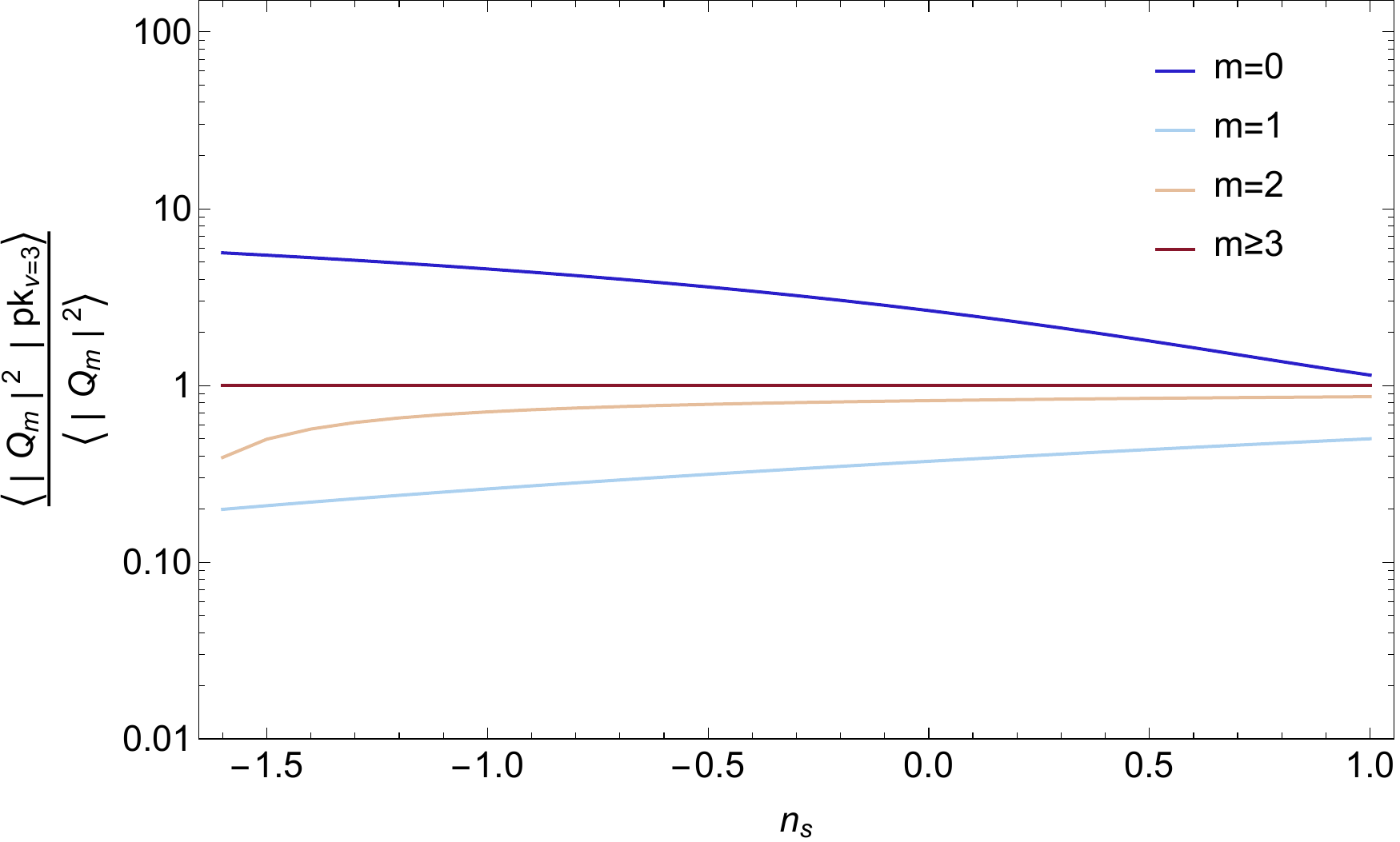} 
\caption{Same as Fig.~\ref{fig:Qm} for central peaks of height $\nu=3$ and separation $r=1$ (in units of the smoothing length) as a function of the spectral index $n_{s}$. 
\label{fig:ns}}
\end{figure}

In this work, we have shown results for a  {power-law} power spectrum ($n_{s}=0$) but the qualitative conclusions can be shown to be almost independent from the spectral index. To illustrate this property, we have computed the multipoles for different slopes of the power spectrum from -1.5 (close to the effective spectral index of the convergence field at cluster scale) to 1 as displayed on Fig.~\ref{fig:ns}. The correction to the monopole and dipole are quasi-linearly suppressed when $n_{s}$ increases while the quadruple is constant for a wide range of slopes $n_{s}\lesssim 1$ and shows only a decrease at very low spectral indices. Overall, it shows that the qualitative picture described in this paper does not depend significantly on the slope of the power spectrum. Investigating the effect of the running is left for future works as no analytical results can be obtained in this case. The study of a more realistic $\Lambda$CDM power spectrum in the non-linear regime will be presented elsewhere.

\section{Comparison with direct measurements in GRF}
\label{sec:GRF}

\begin{figure*}
\centering
\includegraphics[width=.98\columnwidth]{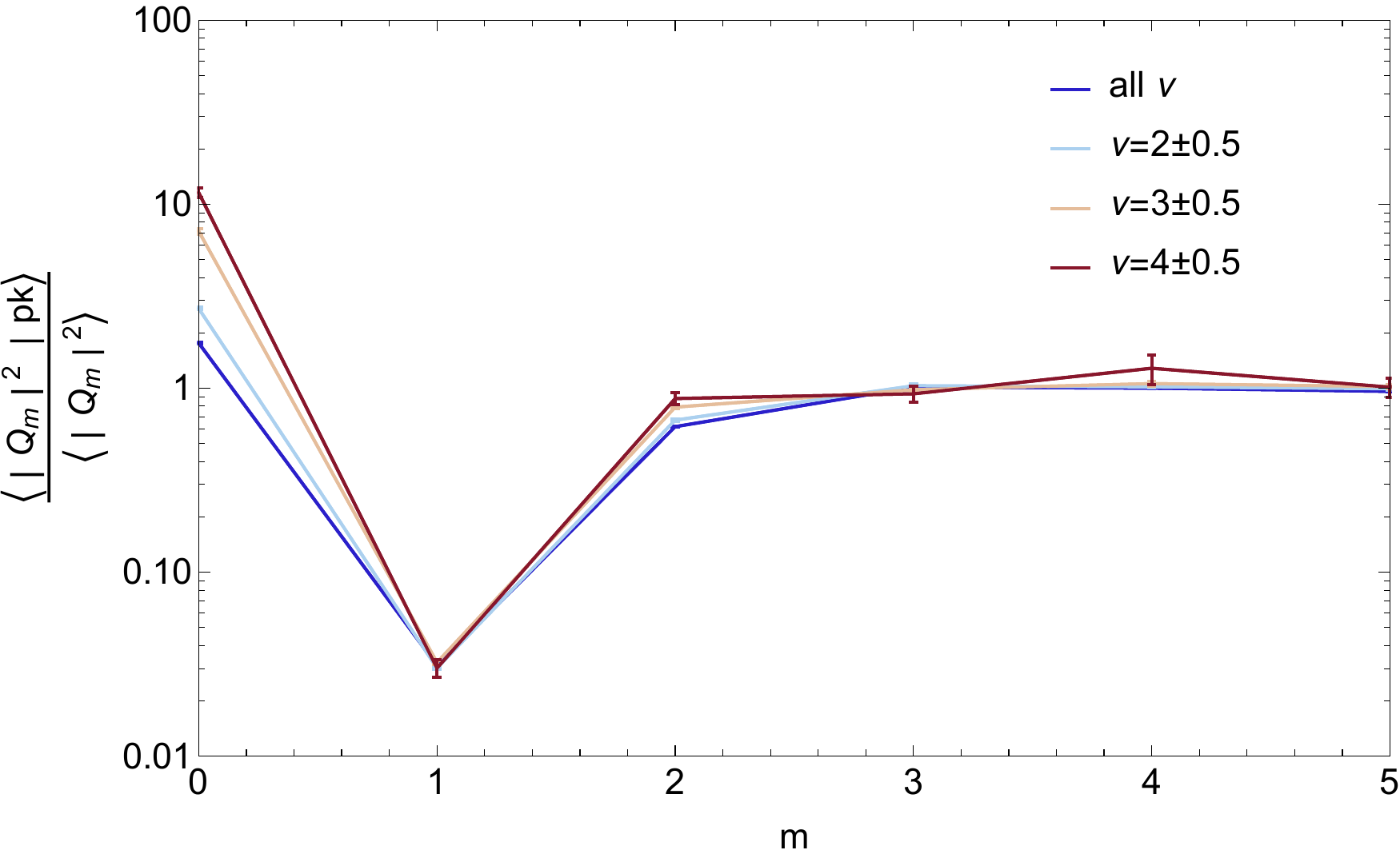} \hskip 0.5cm
\includegraphics[width=.98\columnwidth]{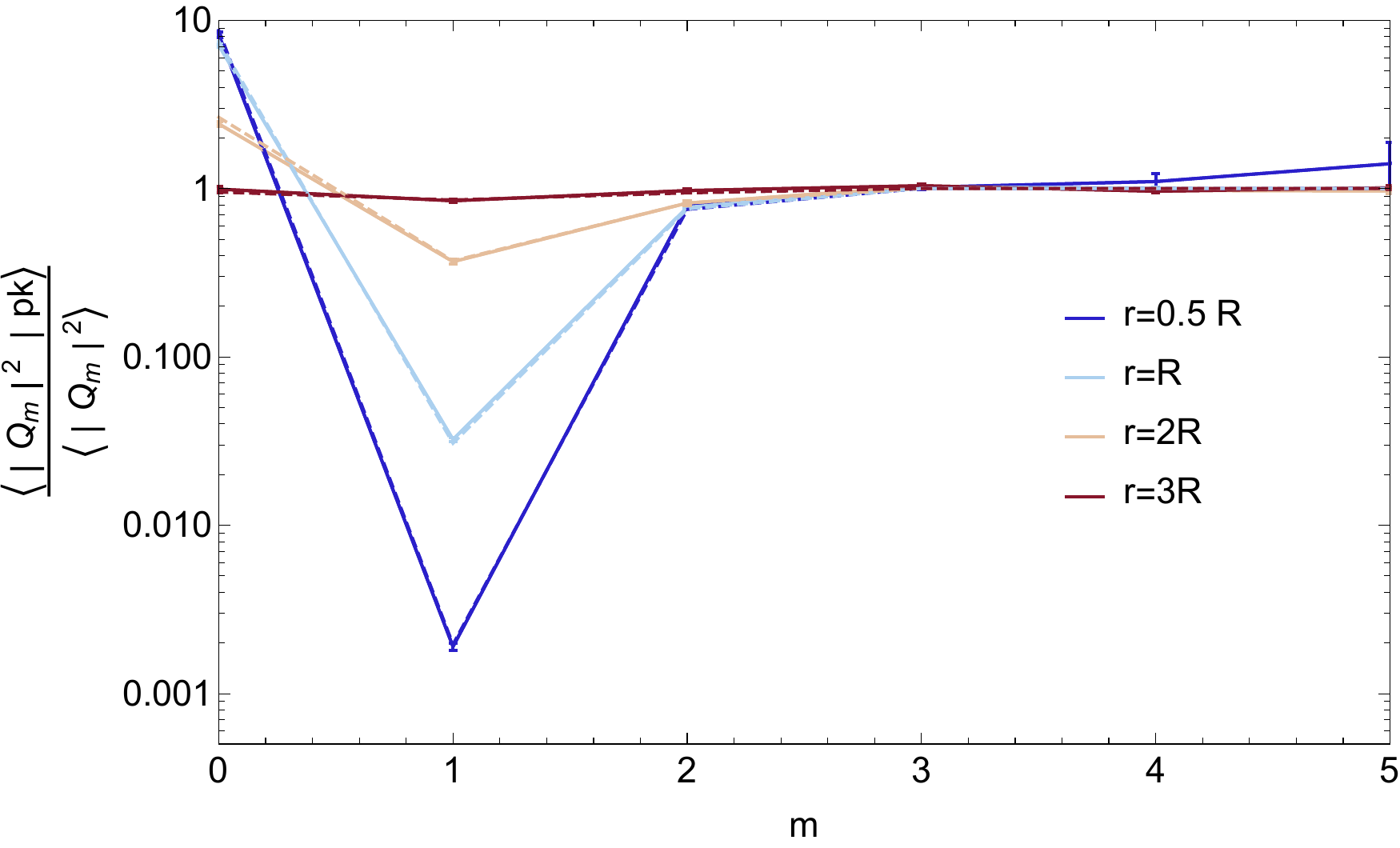} 
\caption{Left-hand panel: Same as the bottom right-hand panel of Fig.~\ref{fig:Qm} for measurements in ten realizations of a $2048^{2}$ 2D GRF smoothed with a Gaussian filter on 8 pixels. The height of the peaks are binned as labelled and the separation considered here is $r=R=8$ pixels. Right-hand panel: Same as left-hand panel when we vary the separation $r$ instead of the peak height which is set to $\nu=3$ here. 
We overplotted the theoretical predictions with dashed lines that are almost indistinguishable from the measurements.
\label{fig:GRF}}
\end{figure*}

Let us  generate ten maps of a $2048^{2}$ GRF with power spectrum $P(k)\propto k^{0}$. Each map is then smoothed with a Gaussian kernel on $R=8$ pixels. A portion of such a map is displayed in the right-hand panel of Fig.~\ref{fig:frame}.

Peaks are then found using the code {\tt map2ext} \citep{2000PhRvL..85.5515C,pogo11}: for every pixel a segment of quadratic surface is fit in the tangent plane based on the field values at the pixel of origin and its neighbours. The position of the extremum of this quadratic surface, its height and its Hessian are computed. The extremum is counted into the tally of the type determined by its Hessian (two negative eigenvalues for peaks) if its position falls within the original pixel. 
Several additional checks are performed to preclude registering extrema in the neighbouring pixels and minimize missing extrema due to jumps in the fit parameters as region shifts to the next pixel. This procedure performs with better than 1\% accuracy when the map is smoothed with a Gaussian filter whose full width at half maximum exceeds 6 pixels.

The field is then interpolated at 100 equally spaced points on the circle located at $r=R$ around each peak and Fourier transformed. Only the square modulus of the Fourier coefficients are stored. For comparison, a similar procedure is followed to estimate the multipolar decomposition around the same number of random points in the field.

The resulting multipolar decomposition measured in GRF is displayed on Fig.~\ref{fig:GRF} for various peak heights and separations. Those measurements are in very good agreement with the theoretical predictions described in Sect.~\ref{sec:multipoles}. The high-$\nu$ approximation used to derive the prediction is therefore shown to be very accurate in the regime $\nu\geq 2.5$. Below this threshold, some departures -- in particular in the quadrupole -- are seen and would require a numerical integration of the equation with the correct peak curvature constraints.

\section{Effect of substructures}
\label{sec:sub}
In practice, measurements in simulations and observations of the angular distribution of the convergence field around clusters naturally involve two separate scales : the (relatively large) scale of the cluster and the (smaller) scale of the convergence field (or the dark matter density field in a N-body simulation) around it. Even if those scales are not identical, they are necessarily highly correlated and the effect described in this paper should persist. To study the effect of substructures,  let us redo the analysis but introducing two different smoothing lengths, one $R_{1}$ for the field $z$ at the location of the peak and one $R_{2}$ at the location of the annulus. In this section only, we will denote $R=R_{2}/R_{1}\leq1$ the corresponding (dimensionless) ratio. The same formalism as described above applies but all the coefficients of the covariance matrix are changed.
Let us first redefine the random variables as
\begin{align} 
z&=\frac{1}{\sigma_0} \kappa({\bf r}_{z}), 
& z_i&=\frac{1}{\sigma_1} \nabla_i \kappa({\bf r}_{z}), 
 &z_{ij}& =\frac{1}{\sigma_2} \nabla_i \nabla_j \kappa({\bf r}_{z}),\\
x&=\frac{1}{\sigma_0} \kappa({\bf r}_{x}),
&y&=\frac{1}{\sigma_0} \kappa({\bf r}_{y}),
\end{align}
where the factors $\sigma_{i}$ are the respective variances of the field, gradient and Laplacian smoothed on scale $R_{1}$.
With this definition, one can easily recompute the coefficients of the covariance matrix. For instance, 
\begin{eqnarray}
&&\left\langle xz \right\rangle=\left\langle yz \right\rangle=\beta^{-n_{s}-2}\, _1F_1\left(\frac{n_{s}}{2}+1;1;-\frac{r^2}{4\beta^{2}}\right),\\
&&\left\langle xy \right\rangle=R^{-\frac {n_{s}+2} 2}\, _1F_1\left(\frac{n_{s}}{2}+1;1;-\frac{r^2}{2R^{2}}(1-\cos\psi)\right),
\end{eqnarray}
\begin{eqnarray}
&&\left\langle xz_{22} \right\rangle=-\frac{\gamma}{2} \beta^{-n_{s}-4}\left[2 \sin ^2(\theta +\psi ) \, _1F_1\left(\frac{n_{s}}{2}+2;1;-\frac{r^2}{4\beta^{2}}\right)\right.\nonumber\\
&&\hskip 2.cm \left.+\cos (2 (\theta +\psi )) \,
   _1F_1\left(\frac{n_{s}}{2}+2;2;-\frac{r^2}{4\beta^{2}}\right)\right],
\end{eqnarray}
where the separation $r$ is again a dimensionless quantity (expressed in units of $R_{1}$) and $\beta$ is the dimensionless quadratic mean of the two smoothing lengths \begin{equation}
\beta=\sqrt{(1+R^{2})/2},
\end{equation}
which ranges from $1/\sqrt2$ (when $R_{2}$ goes to zero) to 1 (when the two smoothing scales are equal $R_{2}=R_{1}$).

An analytical solution for the mean amplitude of the multipoles of the field around a central peak can again be computed. 
For the same example of a  {power-law power spectrum} $P(k)\propto k^{0}$, those multipoles read
\begin{eqnarray}
\left\langle |Q_{0}|^{2}|\textrm{pk}\right\rangle&=&\left\langle |Q_{0}|^{2}\right\rangle+\frac{ {a_{0}^{\rm pk}}}{8\beta^{12}(\nu^{2}-1)}\exp\left(-\frac {r^{2}}{4\beta^{2}}\right)\,,\\
\left\langle |Q_{1}|^{2}|\textrm{pk}\right\rangle&=&\left\langle |Q_{1}|^{2}\right\rangle-\frac{ r^{2}}{4\beta^{8}}\exp\left(-\frac {r^{2}}{4\beta^{2}}\right)\,,
\end{eqnarray}
\begin{eqnarray}
\left\langle |Q_{2}|^{2}|\textrm{pk}\right\rangle&=&\left\langle |Q_{2}|^{2}\right\rangle-\frac{r^{4}}{16(\nu^{2}-1)\beta^{12}}\exp\left(-\frac {r^{2}}{4\beta^{2}}\right)\,,\\
\left\langle |Q_{m}|^{2}|\textrm{pk}\right\rangle&\overset{ m\geq 3}{=}&\left\langle |Q_{m }|^{2}\right\rangle\equiv\frac 1 {R^{2}}\exp\left(-\frac {r^{2}}{2R^{2}}\right)I_{m}\left(\frac{r^{2}}{2R^{2}}\right) \,.
\end{eqnarray}
where $I_{m}$ are the modified Bessel functions of the first kind and
\begin{multline}
 {a_{0}^{\rm pk}}=r^{4}-8\beta^{2}\left[1+\beta^{2}(\nu^{2}-1)\right]r^{2}\\
+8\beta^{4}\left[2+4\beta^{2}(\nu^{2}-1)+\beta^{4}(\nu^{4}-6\nu^{2}+3)\right].\nonumber
\end{multline}
The limit $R=\beta=1$ trivially reduces to the former Eqs.~(\ref{eq:Q0th}-\ref{eq:Qmth}).

\begin{figure}
\centering
\includegraphics[width=\columnwidth]{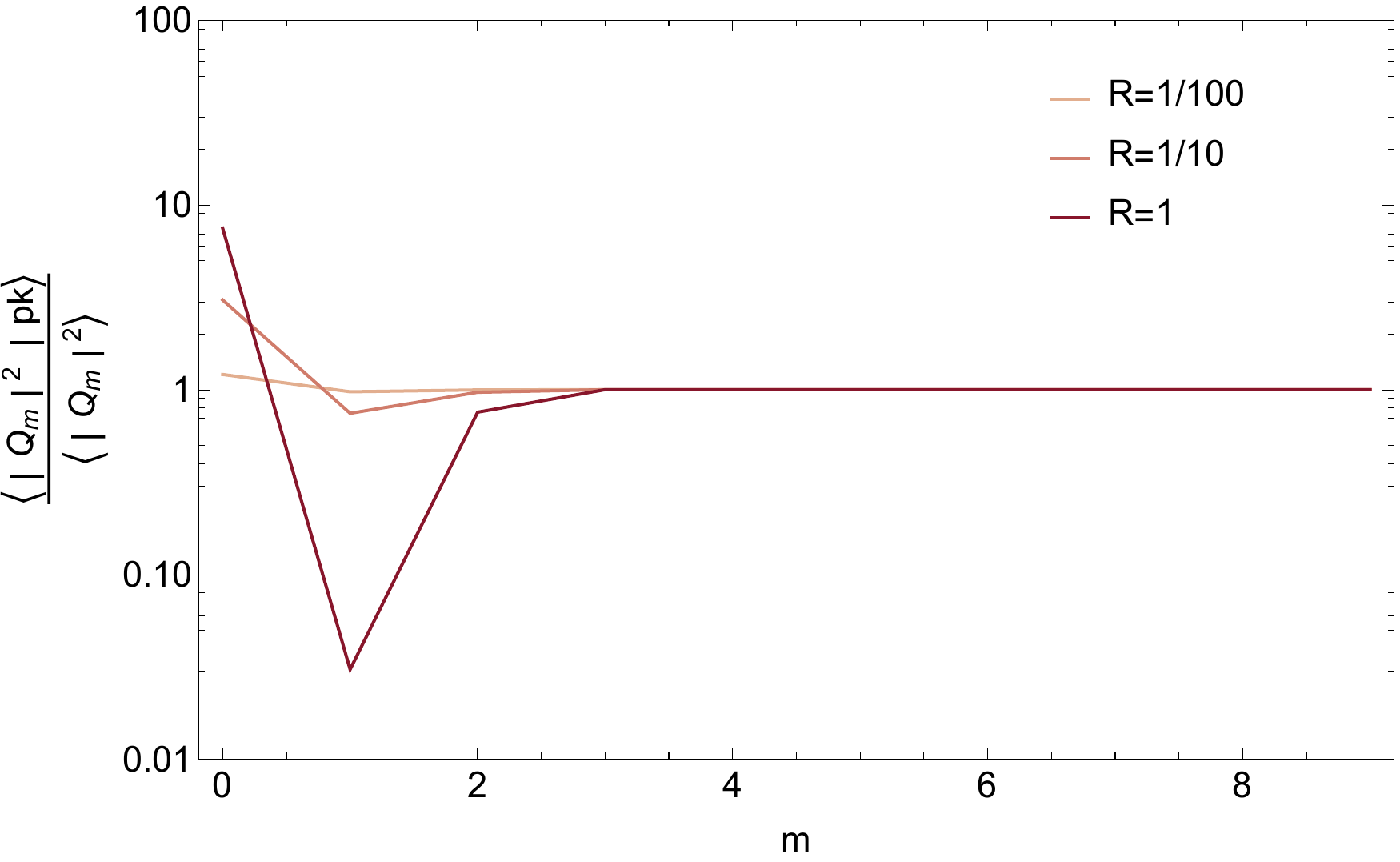} 
\caption{Same as the bottom right-hand panel of Fig.~\ref{fig:Qm} when the field is smoothed at two different scales whose ratio $R$ goes from 1/100 to 1. 
\label{fig:Qm-multi}}
\end{figure}

Those small-scale multipoles are displayed in Fig.~\ref{fig:Qm-multi} for $R=1/100$ to $R=1$. The multi-scale approach described in this section does not modify the $m>2$ multipoles. As expected, the correction due to the peak decreases when $R$ goes to 0 as the scales decorrelate. In addition, we expect that non-linearities and corrections beyond the Hessian will change the power of higher order multipoles.

\section{Beyond the thin shell approximation}
\label{sec:wr}
\begin{figure}
\centering
\includegraphics[width=\columnwidth]{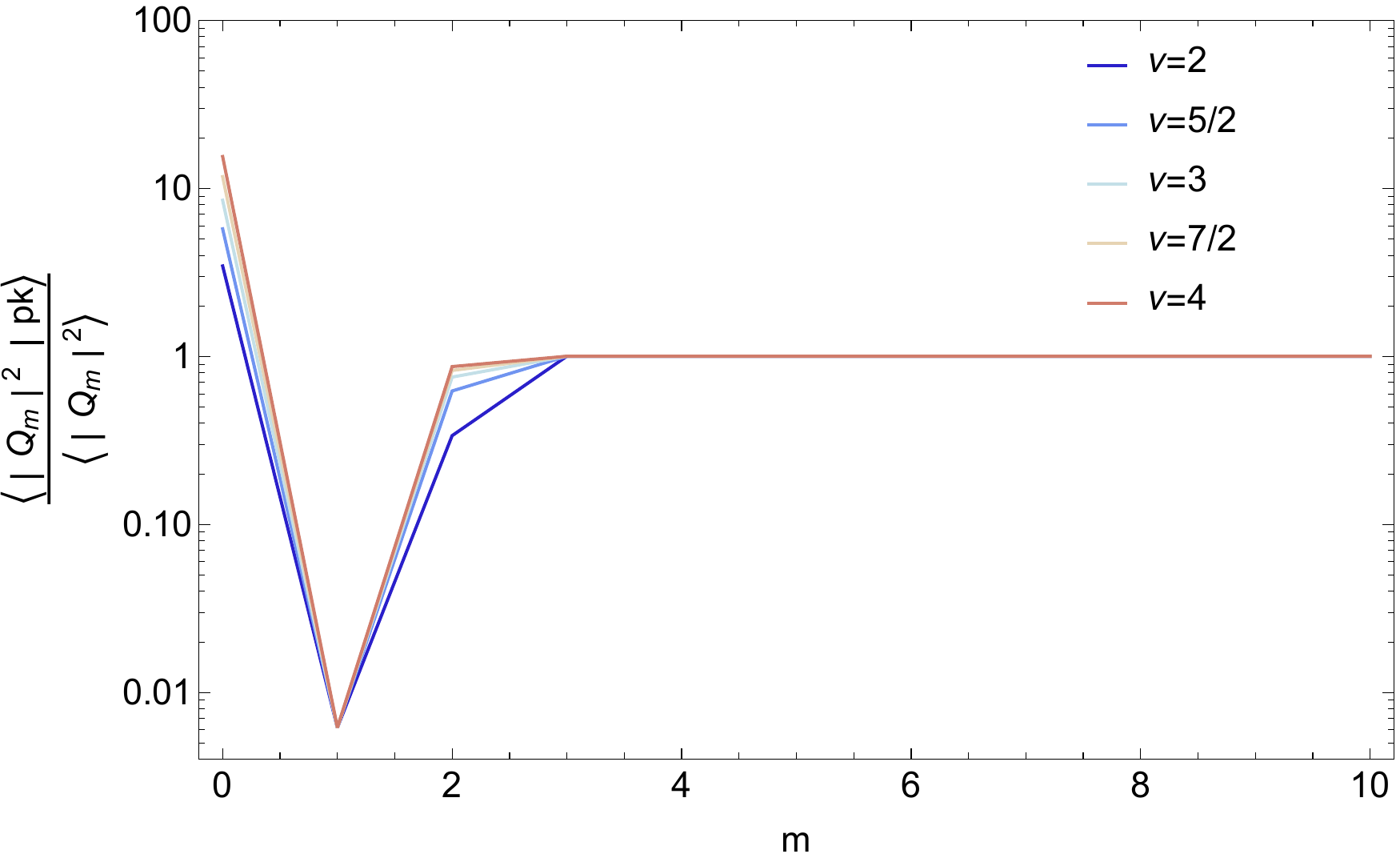} 
\caption{Same as the bottom right-hand panel of Fig.~\ref{fig:Qm} when we apply a radial weight function defined by Eq.~(\ref{eq:weightSB7}) with $R_{\rm max}$ equals the smoothing length and $\alpha=0.5$ (which means that the minimum radius considered is half the smoothing length). 
\label{fig:Qm-radial}}
\end{figure}

The effect of the radial weight function in Eq.~(\ref{eq:def-multipoles}) can be studied by relaxing the assumption that ${\bf r}_{x}$ and ${\bf r}_{y}$ are on a same infinitely thin shell around the central peak in ${\bf r}_{z}$. Let us therefore consider the general setting for which ${\bf r}_{x}$ is at a distance $r$ from the central peak and ${\bf r}_{y}$ at a distance $r'$.
In this case, the constrained two-point correlation function reads
\begin{eqnarray}
\label{eq:r1r2}
\frac{\left\langle \kappa(r,\theta) \kappa(r,\theta+\psi)|{\rm pk}(\nu)\right \rangle}{\sigma_{0}^{2}}\!\!&=&\!\!{  \xi(|{\bf r}_{x}-{\bf r}_{y}|)}\!+\!\!\frac{2f_{21}f_{21}'}{\nu^{2}\!-\!1}\!\!+\!2(f_{11}'f_{21}\!+\!f_{11}f_{21}')\nonumber\\
&+&\!\!\!\frac{\nu^{4}\!-\!6\nu^{2}\!+\!3}{\nu^{2}\!-\!1}f_{11}f_{11}'\!\!-\!\frac{n_{s}\!\!+\!2}{4}r^{2}\!\cos\psi f_{22}f_{22}'\nonumber\\
 &-&\!\!\!\frac{2\cos(2\psi)}{\nu^{2}\!-\!1}(f_{21}\!-\!f_{11})(f_{21}'\!-\!f_{11}'),
\end{eqnarray}
where $f$, $f'$ and the unconstrained correlation function { $ \xi(|{\bf r}_{x}-{\bf r}_{y}|)$} are functions of the following Kummer confluent hypergeometric functions
\begin{eqnarray}
&&f_{ij}=\pFq{1}{1}{\frac {n_{s}} 2 +i}{j}{-\frac{r^{2}}{4}},\\
&&f_{ij}'=\pFq{1}{1}{\frac {n_{s}} 2 +i}{j}{-\frac{r'^{2}}{4}},\\
&&{  \xi(|{\bf r}_{x}-{\bf r}_{y}|)=\pFq{1}{1}{\frac {n_{s}} 2 +i}{j}{-\frac{r^{2}+r'^{2}-2rr'\cos\psi}{4}}}.
\end{eqnarray}
It can easily be checked that Eq.~(\ref{eq:r1r2}) trivially reduces to Eq.~(\ref{eq:kk-pk}) when $r'=r$.
As an illustration, for a  {power-law} power spectrum $P(k)\propto k^{0}$, the correction to the unconstrained correlation function {  $\xi(|{\bf r}_{x}-{\bf r}_{y}|)$} reads
\begin{equation}
\frac{\left\langle \kappa(r,\theta) \kappa(r',\theta+\psi)|{\rm pk}\right \rangle}{\sigma_{0}^{2}}-{  \xi(|{\bf r}_{x}-{\bf r}_{y}|)}=
\exp\left(-\frac{r^2+r'^{2}}{4}\right)\delta\xi\,,
\end{equation}
with
\begin{equation}
\delta\xi\!=\!\frac{\!8 (\nu^{2}\!-\!1) ^2\!-\!4
   \nu ^2 (r^2\!+\!r'^{2}) \!+\!r^2r'^{2}(1\!-\!\cos 2 \psi)\!-\!4 \left(\nu ^2\!-\!1\right) rr' \!\cos \psi \! }{8 \left(\nu ^2\!-\!1\right)}.\nonumber
   \end{equation}
From Eq.~(\ref{eq:r1r2}), one can now easily compute the statistics of the multipoles including the radial weight function $w_n(r)$ that appears in Eq.~(\ref{eq:def-multipoles}). We find again that only the amplitude of the first three multipoles are affected by the peak constraint 
\begin{eqnarray}
\left\langle |Q_{0}|^{2}|\textrm{pk}\right\rangle&\!\!\!=\!\!\!\!\!&\left\langle |Q_{0}|^{2}\right\rangle+\!\!\int_{{\cal S}_{0}}\!\!\!\frac{r^{4}\!-\!8\nu^{2}r^{2}\!+\!8(\nu^{2}\!-\!1)^{2}\!}{8(\nu^{2}\!-\!1)}\exp\left(-\frac {r^{2}}{2}\right)\label{eq:Q0th-rad},\\
\left\langle |Q_{1}|^{2}|\textrm{pk}\right\rangle&\!\!\!=\!\!\!&\left\langle |Q_{1}|^{2}\right\rangle-\!\!\int_{{\cal S}_{1}}\!\frac {r^{2}}4 \exp\left(-\frac {r^{2}}{2}\right),\\
\left\langle |Q_{2}|^{2}|\textrm{pk}\right\rangle&\!\!\!=\!\!\!&\left\langle |Q_{2}|^{2}\right\rangle-\!\!\int_{{\cal S}_{2}}\!\frac 1 {16}\frac{r^{4}}{\nu^{2}\!-\!1}\exp\left(-\frac {r^{2}}{2}\right),\\
\left\langle |Q_{m}|^{2}|\textrm{pk}\right\rangle&\!\!\!\overset{ m\geq 3}{=}\!\!\!&\left\langle |Q_{m }|^{2}\right\rangle
\label{eq:Qmth-rad},
\end{eqnarray}
where $\int_{{\cal S}_{m}}$ stands for the following 2D radial integral
\begin{equation}
\int_{{\cal S}_{m}} f(r,r')=(2\pi)^{2}\sigma_{0}^{2}\int r\der r\, r'\der r' \,  r^{n} w_n(r)  r'^{m} w_m(r')\,.
\end{equation}
Fig.~\ref{fig:Qm-radial} shows the resulting multipoles for a radial weight function defined following \cite{SB97} as
\begin{equation}\label{eq:weightSB7}
R_{\rm max}^{1+m} \, w_m(r)  = \frac{1}{x^{1+m}+\alpha^{1+m}} - \frac{1}{1+\alpha^{1+m}}   + \frac{(1+m)(x-1)}{(1+\alpha^{1+m})^2}\,
\end{equation}
over the range $x=r/R_{\rm max} \in [\alpha, 1]$ and zero elsewhere (which was found to be optimal for an isothermal mass distribution). The qualitative picture does not change : the most affected multipole is the dipole whose power is significantly reduced by the peak constraint, the monopole and quadrupole are slightly affected in a $\nu$-dependant way and all other coefficients are unaffected.

\begin{figure*}
\centering
\includegraphics[width=0.8\columnwidth]{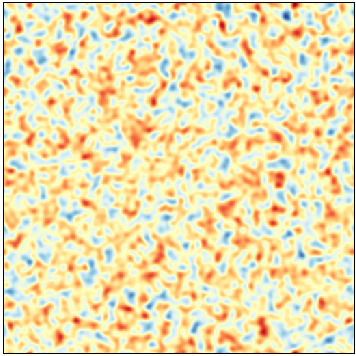}
\includegraphics[width=0.8\columnwidth]{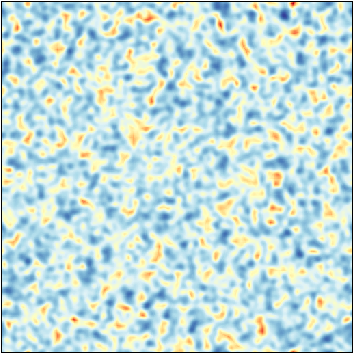} 
\includegraphics[width=0.8\columnwidth]{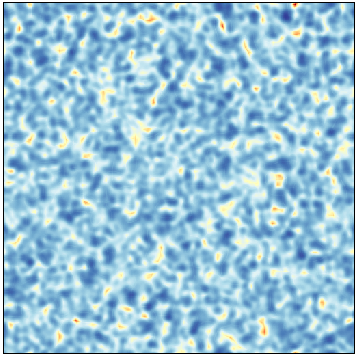} 
\includegraphics[width=0.8\columnwidth]{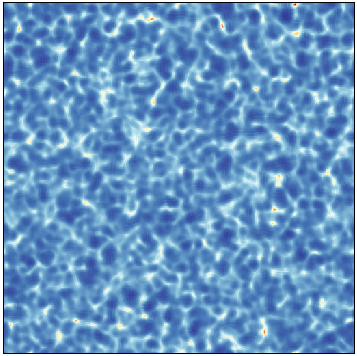} 
\caption{{ 
A slice through the fields $n_{\rm step}=$0,1,2 and 3 obtained by a Zeldovich displacement of an initial GRF. We measure the multipolar moments around all points of height $\nu>2$ in those maps.
\label{fig:zeldoGRF}}}
\end{figure*}

{ 
\section{A non-linear theory of harmonic power spectra}
\label{sec:WNL}

In this section, we study the weakly non-linear evolution of the multipolar moments.
We therefore no longer assume that the PDF is Gaussian ${\cal P}(x,y,z)={G}(x,y,z)$. Instead, we expand the PDF around a Gaussian by means of the so-called Gram-Charlier expansion \citep{Cramer46,2009PhRvD..80h1301P}. For simplicity, we will restrict ourselves to the case where we only impose the height of the cluster but not the rest of the peak condition (no zero gradient or constraint on the eigenvalues of the Hessian). We will show that this effect dominates the high multipoles.

\subsection{The Gram Charlier expansion}
The Gaussian PDF has zero means and covariance matrix
\begin{equation}
\mathbf{C}=\left(
\begin{array}{cccccc}
 1 & b&a  \\
b & 1 &a \\
a & a&1 
\end{array}
\right),
\end{equation}
with $a=\xi(r)$ and $b=\xi(2r\sin(\psi/2))$.

We first diagonalise this matrix and use a new set of variables $(u,v,z)$
where 
\begin{align}
w&=\frac{y-a z}{\sqrt{1-a^{2}}}\\
u&=\frac{x(1-a^{2})+y(a^{2}-b)+z a (b-1)}{\sqrt{(1-b)(1-2a^{2}+b)(1-a^{2})}}
\end{align}
so that
\begin{equation}
{\cal P}(u,w,z)=N(u)N(w)N(z)
\end{equation}
with $N$ a normal distribution of zero mean and unit variance.

Following \cite{gay12,rds}, we then use a Gram-Charlier expansion of the PDF
\begin{equation}
{\cal P}(u,w,z)={G}(u,w,z)\left[1+\!\!\!\sum_{i+j+k=3}^{\infty}\frac{H_{i}(u)H_{j}(w)H_{k}(z)}{i!\,j!\,k!}\left\langle u^{i} w^{j} z^{k}\right\rangle_{\rm GC}\right]\nonumber
\end{equation}
where $H_{i}$ represent probabilistic Hermite polynomials and the Gram-Charlier coefficients are given by
\begin{equation}
\left\langle u^{i} w^{j} z^{k}\right\rangle_{\rm GC}\!\!\!=\!\left\langle H_{i}(u)H_{j}(w)H_{k}(z)\right\rangle.
\end{equation}
Once the joint PDF is known, we can compute the annulus two-point correlation function $\left\langle xy |z=\nu\right\rangle$ as
\begin{equation}
\left\langle xy |z=\nu\right\rangle=\frac{\int{\dd x\,\dd y\,\cal P}(u(x,y,z),w(y,z),\nu) x y}{{\cal P}(z=\nu)}\,,
\end{equation}
which can be rewritten
\begin{equation}
\label{eq}
\left\langle xy |z=\nu\right\rangle=\frac{\int{\dd u\,\dd w\,\cal P}(u,w,\nu) x(u,w,\nu) y(u,w,\nu)}{{\cal P}(z=\nu)}\,,
\end{equation}
where $xy$ is a polynomial of $u$ and $w$
\begin{multline}
xy= a^{2} z^{2}+uza\sqrt{\frac{(1-b)(1-2a^{2}+b)}{1-a^{2}}}+w z a \frac{1+b-2a^{2}}{\sqrt{1-a^{2}}}\\
+uw \sqrt{(1-b)(1-2a^{2}+b)}+(b-a^{2})w^{2}
\end{multline}
which is the sum of four terms proportional respectively to $H_{0}(u)H_{0}(w)$,  $H_{1}(u)H_{0}(w)$,  $H_{0}(u)H_{1}(w)$, $H_{1}(u)H_{1}(w)$ and $H_{0}(u)(H_{2}(w)+H_{0}(w))$ as $H_{0}(x)=1$, $H_{1}(x)=x$ and $H_{2}(x)=x^{2}-1$. Using the property of orthogonality of Hermite polynomials, it is then easy to compute Eq.~\ref{eq} so that eventually
\begin{equation}
\left\langle xy |z=\nu\right\rangle=b+a^{2}(\nu^{2}-1)+\Delta_{\rm NL} \,,
\end{equation}
where the non-linear contribution reads
\begin{equation}
\Delta_{\rm NL}\!\!=\!\frac{\sum_{k=1}^{\infty}\!\frac{H_{k}(\nu)}{k!}\!\left\langle H_{k}(z)\!\left[\!2a\nu(x\!-\!az)\!+\!xy\!-\!b\!-\!2azx\!+\!a^{2}z^{2}+a^{2}\right]\right\rangle}{1+\!\sum_{k=3}^{\infty}\frac{H_{k}(\nu)}{k!}\left\langle z^{k}\right\rangle_{\rm GC}},\nonumber
\end{equation}
which, at first order {  in $\sigma_{0}$ -- the amplitude of fluctuations --}, is given by\footnote{  An easy way to get this expression is to only keep Gram-Charlier coefficients $\left\langle u^{i} w^{j} z^{k}\right\rangle_{\rm GC}$ for which $i+j+k=3$ which were shown to be equivalent to cumulants and correspond exactly to the first-order correction, proportional to $\sigma_{0}$ \citep{gay12}.}
\begin{equation}
\Delta_{\rm NL}^{(1)}\!\!=a^{2}(2\nu-\nu^{3})\left\langle z^{3}\right\rangle+a(\nu^{3}-3\nu)\left\langle xz^{2}\right\rangle+\nu\left\langle x y z\right\rangle.\nonumber
\end{equation}
In terms of multipoles, it means that for $m>0$, we get a 
non-linear bias given by
\begin{equation}
\label{eq:pred}
\frac{\left\langle |Q_{m}|^{2}|\nu\right\rangle}{\left\langle |Q_{m}|^{2}\right\rangle}=1+\nu\frac{\left\langle x y z\right\rangle_{m}}{b_{m}}+{\cal O}(\sigma_{0}^{2})\,,
\end{equation}
where the subscript $m$ refers to the associated multipole of order $m$. The multipoles near a high density cluster is therefore biased compared to random locations, this bias being proportional to the height $\nu$ with a proportionality coefficient related to the ratio between the isosceles three-point function $\left\langle x y z\right\rangle$ and the two-point correlation function of its base $b=\left\langle x y \right\rangle=\xi(2r\sin(\psi/2))$. Note that the all-order expression is also easily obtained once it is realised that the only terms which depend on the angle $\psi$ are $b$ and the cumulants involving the product $xy$
\begin{equation}
\frac{\left\langle |Q_{m}|^{2}|\nu\right\rangle}{\left\langle |Q_{m}|^{2}\right\rangle}=1+\!\frac{\sum_{k=1}^{\infty}\!\frac{H_{k}(\nu)}{k!}\!\left\langle H_{k}(z)\!\left(\!xy\!-\!b\right)\right\rangle_{m}}{b_{m}\left(1+\!\sum_{k=3}^{\infty}\frac{H_{k}(\nu)}{k!}\left\langle z^{k}\right\rangle_{\rm GC}\right)}\,.
\end{equation}
The monopole is also easy to compute
\begin{equation}
\frac{\left\langle |Q_{0}|^{2}|\nu\right\rangle}{\left\langle |Q_{0}|^{2}\right\rangle}\!=\!1+\frac{a^{2}(\nu^{2}\!-\!1)}{b_{0}}+\delta q_{0}^{\rm NL}\,,
\end{equation}
where the first order non-linear correction reads
\begin{equation}
\delta q_{0}^{\rm NL}=+\nu\frac{a^{2}(2\!-\!\nu^{2})\left\langle z^{3}\right\rangle+a(\nu^{2}\!-\!3)\left\langle xz^{2}\right\rangle\!+\!\left\langle x y z\right\rangle_{0}}{b_{0}}+{\cal O}(\sigma_{0}^{2}).\nonumber
\end{equation}

\subsection{Comparison with simulations}

To test the $m>0$ prediction, we have generated various GRF and displaced the density field following a Zeldovich displacement with different time steps denoted $n_{\rm step}=$0,1,2 and 3 (from Gaussian to more evolved fields). In practice, we compute the displacement field as the gradient of the gravitational potential by FFT and we multiply by a constant times $n_{\rm step}$. We then move the mass in each pixel according to this displacement and distribute it to the eight closest pixels. Those fields are illustrated on Fig.~\ref{fig:zeldoGRF}.
We measure the multipoles around field point of height $\nu>2$ together with the mean height of those peaks  (resp. $\bar \nu=2.36, 2.52, 2.63, 2.86$) and the multipolar decomposition of the bispectrum $\left\langle x y z\right\rangle_{m}$. The result is displayed on Fig.~\ref{fig:zeldo} and shows a fair agreement of the $m>0$ multipoles with the prediction given in Eq.~\ref{eq:pred}.

\begin{figure}
\centering
\includegraphics[width=0.96\columnwidth]{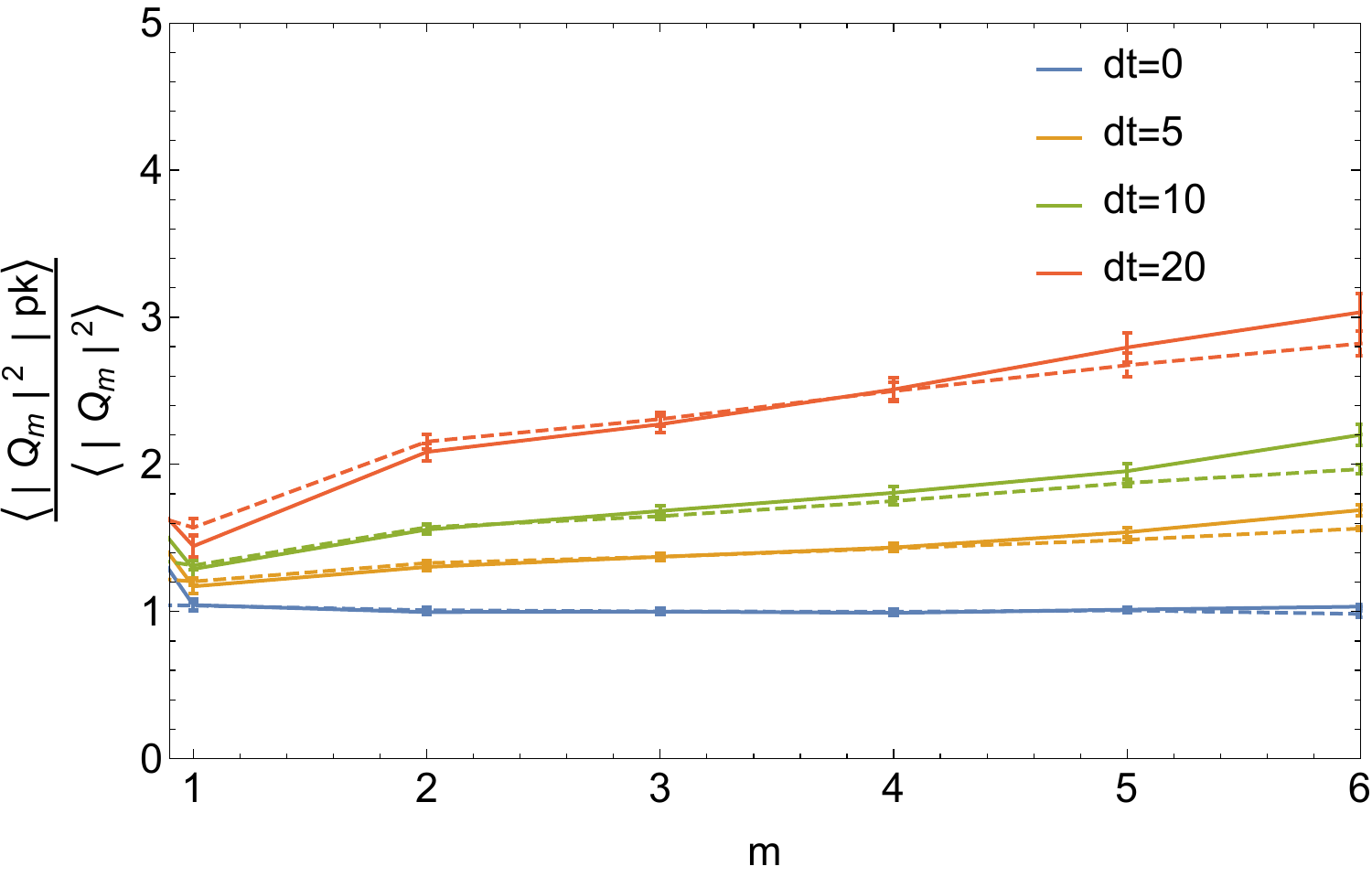} 
\caption{
Ratio between the multipolar moments at a distance $r=8$ pixels from a field point of height $\nu>2$ compared to random locations, for a GRF displaced following the Zeldovich approximation for different time steps between 0 and 20 as labeled and smoothed over 3 pixels. The solid line is the measurements and the dashed line is the prediction for $m>0$ given by Eq.~\ref{eq:pred}. 
\label{fig:zeldo}}
\end{figure}
}

\section{Conclusions}
\label{sec:conclusion}

We have computed the statistics of the multipolar moments around a peak for a generic two-dimensional Gaussian field as a proxy for the azimuthal distribution of matter around clusters seen by weak gravitational lensing experiments. 
For rare enough peaks ($\nu\gtrsim 2.5$), all results are completely analytical. It is shown that only the monopole, dipole and quadrupole are affected by the central peak while higher order multipoles are essentially left unchanged by the peak constraint. Overall, the dominant effect we find is a significant drop in the dipole coefficient as expected from the zero gradient condition. Substructures in the Gaussian field and the addition of a radial weighting function do not change this qualitative picture.

This feature in the dipole can also be detected in numerical simulations of structure formation as will be shown in a forthcoming paper  {\citep{gouin}}. 
We anticipate that higher order corrections will also emerge from the non-linear evolution of the density field in the vicinity of peaks beyond the Gaussian picture described here but also from possible departure from the peak model itself which, as we showed in this paper, boils down to modifying the  power in the monopole, dipole and quadrupole only. 
 {As an illustration, we have computed the non-linear bias of the multipolar moments due to the height of the cluster. This bias is proportional to the height $\nu$ and to the variance of the field $\sigma$ by means of the rescaled bispectrum.}
This approach based on the statistics of multipolar moments in the convergence field around clusters will soon be applied to data (Gavazzi et al, in prep.).

Extensions of this analytical work in the future might include i) an investigation of the accuracy of the large $\nu$ approximation and a precise numerical integration in the regime of intermediate contrasts where this approximation breaks down,
ii) a study of the effect of the scale-dependence of the power spectrum.

\begin{acknowledgements}
This work is partially supported by the grants ANR-13-BS05-0005 of the French Agence Nationale de la Recherche.
This work has made use of the Horizon cluster on which the GRF maps were generated, hosted by the Institut d'Astrophysique de Paris.
We warmly thank D. Pogosyan for insightful discussions, his careful reading of the manuscript and for providing us with his code {\tt map2ext} to detect extrema in 2D maps.
We also thank S. Rouberol for running the Horizon cluster for us and D. Munro for freely distributing his Yorick programming language and opengl interface (available at \url{yorick.sourceforge.net}).
\end{acknowledgements}

\bibliographystyle{aa}
\bibliography{references}
\balance

\appendix

\section{Typical peak geometry}
\label{app:geom}

\begin{figure}
\centering
\includegraphics[width=8cm]{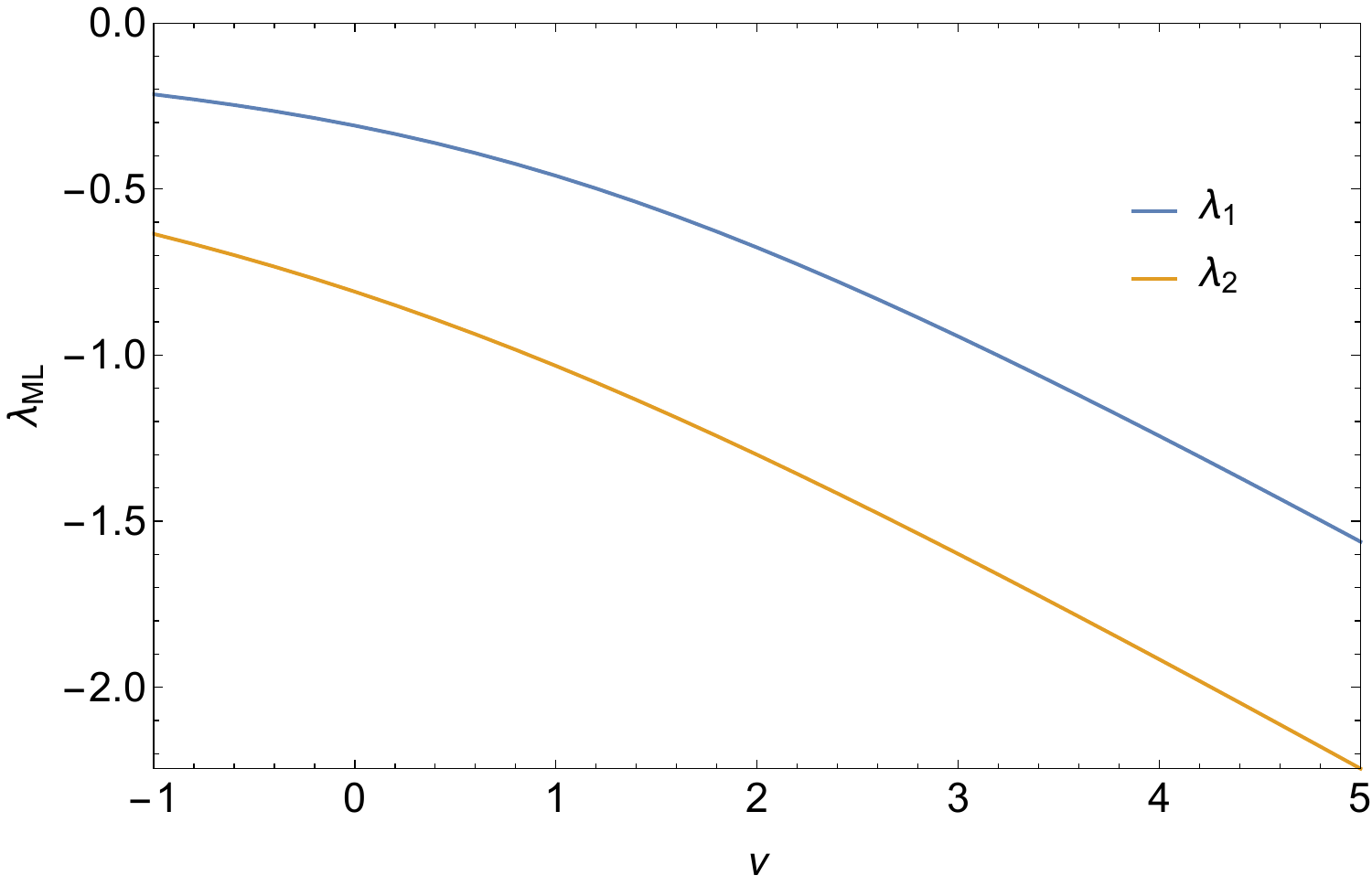}\\
\hskip 0.3 cm\includegraphics[width=7.7cm]{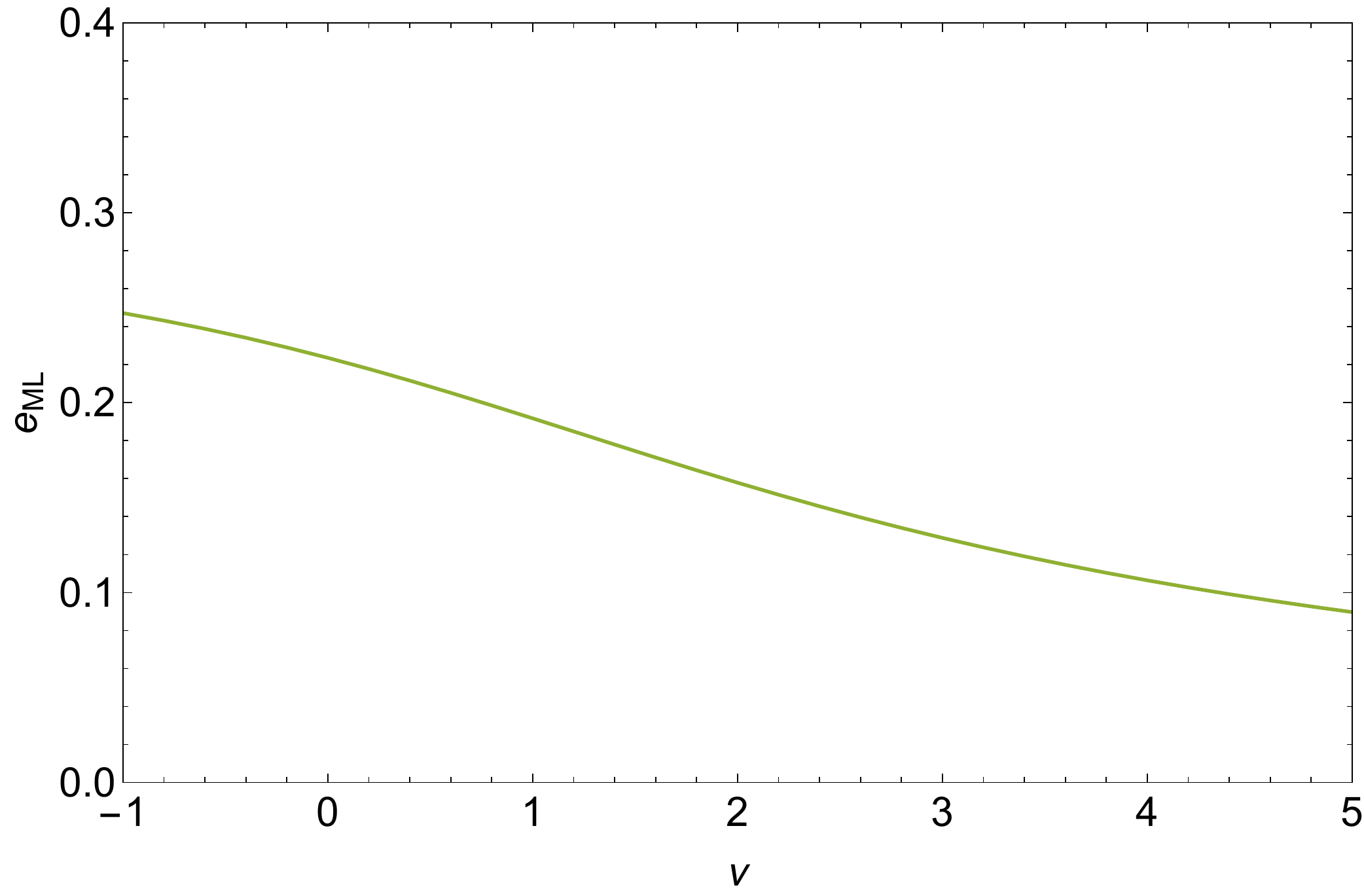} 
\caption{Most likely peak curvatures as a function of the peak height $\nu$ (top panel) and the corresponding most likely peak ellipticity (bottom panel). Those results do not depend on the spectral index. \label{fig:geom-pk}}
\end{figure}

The typical geometry of a Gaussian peak in two dimensions can easily be computed.
Starting from the Gaussian joint PDF of the field value $\nu$ and local curvatures $\lambda_{1}>\lambda_{2}$  \citep{doroshkevich70,pogosyan/pichon/etal:2009}
 \begin{equation}
 {\cal P}(\nu,\lambda_{1},\lambda_{2})\!=\!\!\frac{2\sqrt{J_{2}}}{\pi\sqrt{1-\gamma^{2}}}\exp \left(-\frac 1 2 \!\left(\frac{\nu+\gamma I_{1}}{\sqrt{1-\gamma^{2}}}\right)^{2}\!\!-\!\frac 1 2 I_{1}^{2}\!-\!J_{2}\!\right),
 \end{equation}
 where $I_{1}=\lambda_{1}+\lambda_{2}$ and $J_{2}=(\lambda_{1}-\lambda_{2})^{2}$,
one can show that the PDF for a peak to have height $\nu$ and geometry $0>\lambda_{1}>\lambda_{2}$ reads \citep{BBKS,ATTT}
  \begin{multline}
 {\cal P}(\nu,\lambda_{1},\lambda_{2}|\rm{pk})=\frac{8\sqrt{3}(\lambda_{1}-\lambda_{2})\lambda_{1}\lambda_{2}}{\pi\sqrt{1-\gamma^{2}}}\times
 \\
 \exp \left(-\frac 1 2 \!\left(\frac{\nu+\gamma (\lambda_{1}+\lambda_{2})}{\sqrt{1-\gamma^{2}}}\right)^{2}\!-\!\frac 1 2(\lambda_{1}+\lambda_{2})^{2}\!-\!(\lambda_{1}\!-\!\lambda_{2})^{2}\right)
\,.\label{eq:2Dgeom}
 \end{multline}
 It has to be emphasized that here we do impose exactly the peak constraint given by Eq.~(\ref{eq:constraintpk}).
The most likely value of the peak height and curvatures is therefore given by $ \nu_{\star}=\sqrt{7/3}\,\gamma$, $\lambda_{1,2\star}=(-\sqrt{7/3}\pm\sqrt{1/3})/2$ which corresponds to an ellipticity $e_{\star}=1/(2\sqrt 7)$.

If $\nu$ is fixed e.g to a rare value $\nu_{r}=3$, the maximum of the PDF given by Eq.~(\ref{eq:2Dgeom}) is changed to  $\lambda_{1r}=-0.94$ and $\lambda_{2r}=-1.6$ so that the ellipticity of the peak is given by $e_{r}=0.13$, independently from the spectral parameter $\gamma$.
The evolution of the most likely peak curvatures as a function of height is shown in Fig~\ref{fig:geom-pk}. In particular, it illustrates the well-known result that high peaks are increasingly spherical.

\end{document}